\documentclass[aps,twocolumn,prb,showpacs]{revtex4-1}
\usepackage[latin9]{inputenc}
\usepackage{amsmath}
\usepackage{amssymb}

\makeatletter
\@ifundefined{textcolor}{}
{%
 \definecolor{BLACK}{gray}{0}
 \definecolor{WHITE}{gray}{1}
 \definecolor{RED}{rgb}{1,0,0}
 \definecolor{GREEN}{rgb}{0,1,0}
 \definecolor{BLUE}{rgb}{0,0,1}
 \definecolor{CYAN}{cmyk}{1,0,0,0}
 \definecolor{MAGENTA}{cmyk}{0,1,0,0}
 \definecolor{YELLOW}{cmyk}{0,0,1,0}
}

\usepackage{amsfonts}\usepackage{mathrsfs}\usepackage{epsfig}\usepackage{amsbsy}\usepackage{bm}

\DeclareMathOperator{\Tr}{Tr}
\usepackage{hyperref}
\usepackage{amsmath}
\usepackage{xcolor}
 \usepackage[T1]{fontenc}
\usepackage{amssymb}
\usepackage{natbib}
\usepackage{graphicx}
\usepackage{amstext}
\usepackage{amsfonts}
\usepackage{empheq}
\usepackage{epstopdf}
\usepackage{enumerate}
\usepackage{simplewick}
\DeclareMathOperator{\iim}{Im}
\DeclareMathOperator{\ep}{\epsilon}

\makeatother

\begin{document}

\title{Strongly-Correlated Thermoelectric Transport beyond Linear Response}

\author{Prasenjit Dutt$^{1}$ and Karyn Le Hur$^{2}$}

\affiliation{$^{1}$ Department of Physics, Yale University, New Haven, CT 06520,
USA}

\affiliation{$^{2}$ CPHT, École Polytechnique, CNRS, Palaiseau 91128 Cédex, France}
\begin{abstract}
We investigate nonlinear thermoelectric transport through quantum impurity systems with strong on-site interactions. 
We show that the steady-state transport through interacting quantum impurities in contact with electron reservoirs at significantly different temperatures
can be captured by an effective-equilibrium density matrix, expressed compactly in terms of the Lippmann-Schwinger operators of the system. 
In addition, the reservoirs can be maintained at arbitrary chemical potentials. The interplay between the temperature gradient and bias voltage gives rise to a non-trivial breaking of particle-hole symmetry in the strongly correlated regime, manifest in the Abrikosov-Suhl localized electron resonance. This purely many-body effect, which is in agreement with experimental results, does not arise from mean-field arguments and for weak-interactions. 
\end{abstract}


\maketitle

\section{Introduction}

Non-equilibrium phenomena have become a topic of intense study for theorists and experimentalists alike.\cite{Deblock,Zurek, Boulat, Delattre, Pavel, Altimiras, MoraLeHur, Martin, Werner, Chung, Thomas, Etzioni, Schiro, Latta, Basset, Knap, Gabelli, Camille, Frey,Fukuhara,Cohen,OrthLeHurAdilet,Kennes,Feve,Dubois,Steinberg} In particular, the ability to induce out-of-equilibrium experimental conditions at the nanoscale has led to the observation of unforeseen features where existing equilibrium methods prove insufficient. A widely studied class of systems in this category are quantum impurities, which are nanofabricated using electron-beam lithography on (Al,Ga)As/GaAs heterostructures containing a two-dimensional electron gas (2DEG).\cite{Goldhaber_1998,Delft} The manipulation of thermoelectric transport in these mesoscopic systems presents a plethora of potential future applications, for example nanoengines which are ideal harvesters of heat energy and in magnetic nanostructures (spin-caloritronics).\cite{Jacquod, Lin,Buttiker_ne, Lim} In particular, advances in nanongineered thermoelectric materials aiming to maximize the Seebeck coefficient has extremely promising functionalities.\cite{Handbook} The effect of electron-electron interactions may enhance the Seebeck coefficient;\cite{CostiZlatic} the latter measures the thermovoltage produced across a sample when a thermal gradient is applied. Adjusting the chemical potentials of the leads and introducing a thermal gradient across a quantum dot (QD), it is also possible to drive the system far beyond the linear response regime.\cite{Massimiliano} Pioneering experiments making use of quantum point contacts were conducted,\cite{Molenkamp2} and cold atomic systems are ideal candidates where similar physics can be replicated.\cite{Esslinger} 

In this paper, our primary goal is to theoretically investigate the nonlinear thermoelectric transport through an interacting QD, in order to rigorously analyze experimental results.\cite{Molenkamp} The nonlinear regime has been recently addressed by treating interactions at a mean-field level.\cite{DavidRosa} Additional theoretical works have also been performed in the context of interacting particles but in the linear regime.\cite{Andreev,Kiselev}

Using specially designed gates one can control the electron states of the dot, and drive the system into the Kondo spin-correlated regime. Here, we find the emergence of a very interesting feature, the apparent breaking of particle-hole symmetry in the localized electron's spectral function as evidenced experimentally.\cite{Molenkamp} This phenomenon is a result of a subtle interplay between the thermal and voltage gradients which emerge beyond the linear regime, when substantial interaction effects exist. We shall provide a physical understanding of this phenomenon as well as a rigorous justification based on the reformulation of the steady-state dynamics of the system in terms of an effective-equilibrium density matrix. 

In fact, the steady-state density matrix has a very intuitive form in terms of the Lippmann-Schwinger operators of the system. A related steady-state formulation was first proposed by Hershfield to describe the system with large bias voltages,\cite{Hershfield} and has been applied to the Anderson impurity,\cite{DuttLeHur,Han,Anders} and interacting resonant level model.\cite{Schiller,Doyon,Bernard} Here, we generalize the approach to additionally include substantial thermal gradients. This approach, which is formally equivalent to the Schwinger-Keldysh approach for the evaluation of steady-state observables,\cite{DuttLeHur}  captures the effect of thermal gradients and potential biases in a particularly transparent manner.  The scattering states are highly delocalized and pervade the entire setup. In fact, for strong tunneling the density profile of scattering states originating in the left lead and right lead are indistinguishable. However, the fact that the lead-index remains a good quantum number and is associated
with a unique temperature and chemical potential is non-trivial. We use the Anderson impurity model coupled to noninteracting electron reservoirs (maintained at different chemical potentials and temperatures), to analyze strongly-correlated transport through a Coulomb-blockaded QD. 

The additional source of particle-hole asymmetry in the electron's spectral function emerging from the interplay between bias voltage and thermal gradient has an intuitive interpretation in terms of the Lippmann-Schwinger scattering states. More precisely, the scattering states are described through well-defined Fermi distribution functions. Writing the scattering states in the s-wave and p-wave basis, only the s-wave component hybridizes with the QD. In fact, for symmetric leads the distribution function for the s-wave part depends exclusively on the effective Fermi distribution function \cite{Oguri} $f^{eff} = 1/2(f_1 + f_{-1})$ where the indices $1$ and $-1$ will refer to the left and right leads, respectively. The spectral function on the QD then only depends on the function $f^{eff}(\omega)$, where 
$\omega$ denotes the frequency.  Particle-hole symmetry in the electron's spectral function on the dot then naturally implies that $f^{eff}(-\omega)=1-f^{eff}(\omega)$.  In the presence of either a bias voltage or a thermal gradient particle-hole symmetry is immediately satisfied, since by swapping the two leads, $f_{eff}$ is unchanged. In this case, particle-hole symmetry is not violated in the electron's spectral function on the QD, assuming that the single particle energy levels on the QD  satisfy particle-hole symmetric condition. Now, when considering both a finite bias voltage and a thermal gradient, one can check that this symmetry in $f_{eff}$ is lost in general - a new source of particle-hole symmetry breaking. 

The primary objective of this paper is to rigorously analyze this additional source of particle-hole symmetry breaking and compare our results with experimental results. 

This paper is organized as follows. In Section II we introduce the steady state density matrix for a QD coupled to Fermi-liquid leads whose chemical potentials and temperatures can be individually tuned. The technical details of the derivation are provided in Appendix \ref{derivation_dm}. In Section III we use the effective equilibrium density matrix to evaluate relevant transport observables in terms of the local Green's function of the electrons on the QD. Having established the required calculational tools we analyze the steady state dynamics of the Anderson impurity model in Section IV. We observe a novel breaking of particle-hole symmetry due to a subtle interplay of thermal gradients and potential biases and present our results in Section V. Finally, we summarize our findings in Section VI. Appendices are devoted to technical details for the sake of clarity.

\section{Effective Equilibrium Description} 
The class of systems we investigate consists of two (or more) macroscopic electronic leads which are individually coupled to a QD. The essential feature of the QD is that by virtue of its size, it effectively allows tunneling of only a few electrons to and from the leads per unit time. In the following discussion, we consider the case of only two macroscopic reservoirs and assume that they are described by non-interacting electrons (more precisely quasiparticles), and the dot is described by a single electronic level with spin projections $\sigma=\uparrow, \downarrow$. The Hamiltonian for this system is given by 
\begin{equation}
H=H_{\text{leads}}+H_{\text{dot}}+H_{\text{tunn}},
\label{eq:Hamiltonian}
\end{equation} 
where 
\begin{subequations}
\begin{align}
&H_{\text{leads}}=\sum_{\alpha k \sigma}\epsilon_{\alpha k}c^{\dag}_{\alpha k\sigma}c_{\alpha k\sigma}\\
&H_{\text{dot}}=\sum_{\sigma}\epsilon_{d}d^{\dag}_{\sigma}d_{\sigma}+H_\text{int}\\
&H_{\text{tunn}}=\frac{1}{\sqrt{\Omega}}\sum_{\alpha k \sigma}t_{\alpha k}\left(c^{\dag}_{\alpha k\sigma}d_{\sigma}+\text{h.c.}\right).
\label{eq:Hamiltonian_parts}
\end{align}
\end{subequations}

Here, we have used explicitly the index $\alpha=\pm 1$ to denote the left ($\alpha=1$) and the right ($\alpha=-1$) leads respectively, and $\Omega$ represents the volume of either lead. The form of interactions on and in the vicinity of the dot will be specified later. 

We do not constrain the temperatures or chemical potentials of the lead in any way. Their thermodynamics is described by the inverse temperatures $\beta_{\alpha}=1/(k_{\text{B}}T_{\alpha})$ and $\mu_\alpha$. The potential bias $\Phi=(\mu_1-\mu_{-1})/2$  and the temperature gradient $\Delta T=T_{1}-T_{-1}$ are responsible for driving particle- and energy-currents through the setup. Generalizing the arguments of Refs. \onlinecite{Hershfield,DuttLeHur,Andrei} we show that in steady-state the system is described by the density matrix
\begin{align}
\rho=\prod_{\alpha=\pm 1}\otimes\exp\left[-\beta_\alpha\sum_{k\sigma}\left(\epsilon_{\alpha k}-\frac{\alpha\Phi}{2}\right)\psi^\dagger_{\alpha k\sigma}\psi_{\alpha k\sigma}\right].
\label{eq:density_matrix}
\end{align}
Details of the derivation are included in Appendix \ref{derivation_dm}.
The operators $\psi^\dagger_{\alpha k\sigma}$ denote the scattering operators of the system, which are given by the operator version of the Lippmann-Schwinger (LS) equation
\begin{equation}
\psi^{\dag}_{\alpha k\sigma}=c^{\dag}_{\alpha k\sigma}+\frac{1}{\epsilon_{\alpha k}-{\cal L}+i\eta}{\cal L}_{\text{T}}c^{\dag}_{\alpha k\sigma}.
\label{eq:LS}
\end{equation}  
Here, the action of the Liouvillians ${\cal L}_{\text{T}}$ and ${\cal L}$ on an arbitrary operator ${\cal O}$ is given by ${\cal L}{\cal O}=[H,{\cal O}]$ and ${\cal L}_{\text{T}}{\cal O}=[H_{\text{tunn}},{\cal O}]$ respectively, and $\eta$ denotes the adiabaticity factor.
In terms of the Lippmann-Schwinger states, the system is describable as two independent non-interacting Fermi seas at different chemical potentials and temperatures. 

Note, though the state $\psi^{\dag}_{\alpha k\sigma}$ is highly delocalized and spans the entire system ({\it i.e.} present in regions 
of different chemical potentials and temperatures), it is effectively described by a single temperature and chemical potential. Furthermore, its thermodynamic properties correspond to those of the reservoir from which it originated. 

\section{Thermoelectric transport}
Under the influence of either $\Phi$ or $\Delta T$ electronic transport occurs in the system and is manifested as non-zero particle- and energy-currents. We use the symmetrized expression for particle and energy currents 
\begin{subequations}
\begin{align}
&I(t)=-\frac{e}{2} \frac{d}{dt}(N_{1}(t)-N_{-1}(t)) = \frac{1}{2}(I_1 - I_{-1}) \\
&I_E(t)=-\frac{e}{2} \frac{d}{dt}(E_{1}(t)-E_{-1}(t)) = \frac{1}{2}(I_1^E - I_{-1}^E),
\label{eq:currents}
\end{align}
\end{subequations}
 since the currents in the left and right lead are balanced in steady state. Here, $N_{\alpha}=\sum_{\alpha k \sigma}\langle c^{\dag}_{\alpha k\sigma} c_{\alpha k\sigma}\rangle$ and 
$E_\alpha =\sum_{\alpha k \sigma}\epsilon_{\alpha k}\langle c^{\dag}_{\alpha k\sigma}c_{\alpha k\sigma}\rangle$ denote the number of electrons and total energy associated with lead $\alpha$. We have identified formally the contributions from each lead
separately for the current and energy current. When the coupling of the dot and the leads is confined to a small spatial region, we can approximate $t_{\alpha k}$ to be roughly constant and be given by $t_\alpha$. Using the description of the system in terms of the effective-equilibrium density matrix in Eq. \eqref{eq:density_matrix} it is possible to write the current in terms of a local quantity, namely the spectral function of the dot, 
\begin{subequations}
\begin{align}
&I=\frac{2 e}{\hbar}\frac{\Gamma_1 \Gamma_{-1}}{\Gamma_1+\Gamma_{-1}}\sum_{\alpha \sigma}\alpha \int d\epsilon_k f_{\alpha}(\epsilon_k)A_d(\epsilon_k)\\
&I_E =\frac{2 e}{\hbar}\frac{\Gamma_1 \Gamma_{-1}}{\Gamma_1+\Gamma_{-1}}\sum_{\alpha \sigma}\alpha \int d\epsilon_k \epsilon_k f_{\alpha}(\epsilon_k)A_d(\epsilon_k).
\label{eq:meir_wingreen}
\end{align}
\end{subequations}
Here, $\Gamma_\alpha=\pi t_\alpha^2\nu_\alpha$ denotes the hybridization of the electrons of the dot and the lead $\alpha$, where $\nu_\alpha$ is its density of states. The derivation of the above expressions is given in Appendix \ref{observables}. 
The spectral function 
\begin{equation}
A_d(\epsilon_k)=-\frac{1}{\pi}\Im[G_{d_\sigma d^{\dag}_\sigma}^{\text{ret}}(\epsilon_k)]
\label{eq:spectral_function}
\end{equation}
is evaluated with respect to the effective-equilibrium density matrix in Eq.\ \eqref{eq:density_matrix}. 

It is instructive to note that observables such as the charge and energy- currents in the steady state can be defined in terms of the electron spectral's function only. A similar conclusion is obtained using the Schwinger-Keldysh formalism by using current
conservation in the steady state limit $I_{1}= - I_{-1}$ \cite{CostiZlatic}. 

The retarded Green's function in Fourier space is \cite{DuttLeHur}
\begin{align}
G^{\text{ret}}_{d_{\sigma}d^{\dag}_{\sigma}}&(\omega)=\left\langle \left\{\frac{1}{\omega+{\cal L}+i\eta}d_{\sigma},d^{\dag}_{\sigma}\right\}\right\rangle\notag\\
&=(\Tr\left[\rho\right])^{-1}\Tr\left[\rho \left\{\frac{1}{\omega+{\cal L}+i\eta}d_{\sigma},d^{\dag}_{\sigma}\right\}\right].
\label{eq:retarded}
\end{align}
Note, it is possible to obtain the same expressions via the Schwinger-Keldysh approach though in that case evaluations should be made on the Keldysh contour with the initial density matrix of the system given by \cite{Meir_Wingreen}

\begin{equation}
\rho_0=\exp\left[-\sum_{\alpha k\sigma}\beta_\alpha\left(\epsilon_{\alpha k}-\frac{\alpha\Phi}{2}\right)c^\dagger_{\alpha k\sigma}c_{\alpha k\sigma}\right].
\label{eq:}
\end{equation}
For simplicity we assume that the couplings to the left and right leads are identical in the ensuing discussion.

The spectral function of the dot is, in general, a function of $\Phi$, $\Delta T$ and the mean temperature $\bar{T}=1/2\sum_{\alpha=\pm 1}T_{\alpha}$. For the simple case of the noninteracting level it has a Lorentzian (bias and temperature independent) form 
\begin{equation}
A^{(0)}_d(\epsilon_k)=\frac{2\Gamma}{\pi}\frac{1}{(\epsilon_k-\epsilon_d)^2+\Gamma^2}.
\label{eq:spectral_function_ni}
\end{equation}
Here, we have defined the total hybridization by $\Gamma\equiv\sum_{\alpha}\Gamma_\alpha$. The general expression for the thermopower is given by 
\begin{equation}
S=-\frac{1}{eT}\frac{\int_{-\infty}^{\infty}d\epsilon_k A_d(\epsilon_k)\left(\epsilon_k -\mu\right) \partial_{\epsilon_k} f(\epsilon_{k})}{\int_{-\infty}^{\infty}d\epsilon_k A_d(\epsilon_k)\partial_{\epsilon_k} f(\epsilon_{k})}.
\label{eq:thermopower}
\end{equation} 
In the absence of interactions this corresponds to the Cutler-Mott formula, which gives an universal value $S\rightarrow\epsilon_d/(e T)$ in the limit of high temperatures, {\it i.e.} $k_{\text B} T\gg \Gamma$.\cite{Landauer,Buttiker}

For the non-interacting case Eq.\ \eqref{eq:LS} can be easily solved and the $\psi_{\alpha k\sigma}$ (which we will denote by $\psi^{(0)}_{\alpha k\sigma}$ in this case) states bear a linear relationship with the $c_{\alpha k\sigma}$ and $d_{\sigma}$ states. This is not the case for an interacting theory, and so we use this simple relationship \cite{DuttLeHur}
\begin{subequations}
\begin{align}
&d_{\sigma}^{\dag}=\frac{t}{\sqrt{\Omega}}\sum_{\alpha k}g^{\ast}_{d}(\epsilon_{k})\psi^{(0)\dag}_{\alpha k\sigma}\\
&c^\dagger_{\alpha k\sigma}=\psi^{(0)\dagger}_{\alpha k\sigma}-
\frac{t^{2}}{\Omega}\sum_{\alpha'k'}\frac{g^{\ast}_{d}(\epsilon_{k'})}{\epsilon_{k}-\epsilon_{k'}+i\eta}\psi^{(0)\dag}_{\alpha'k'\sigma}
\end{align}
\label{eq:inversion}
\end{subequations}
as the basis for perturbative computations. Here, we have defined 

\begin{equation}
g_d(\epsilon_k)\equiv \frac{1}{\epsilon_k -\epsilon_d+i\Gamma}.
\label{eq:gd_def}
\end{equation}

\section{Strongly correlated dynamics} 

The effect of interactions on the transport properties of the system is manifested via $A_d(\epsilon_k)$. 

We consider the Kondo spin-correlated quantum dot whose dynamics can be effectively described by the Anderson impurity model.\cite{Anderson_1961, Molenkamp, Hewson} The Coulomb interactions on the dot are given by the Hamiltonian 
\begin{equation}
H_\text{int}=\frac{U}{2} \hat{n}_d(\hat{n}_d-1),
\label{eq:interactions_anderson}
\end{equation}
where $\hat{n}_d=\sum_{\sigma}\langle d^{\dag}_\sigma d_\sigma\rangle$. We analyze the system by fixing $\epsilon_d=-U/2+\Delta$. The case $\Delta=0$ is referred to as the particle-hole symmetric point \cite{Hewson}; and the extent of deviations from it are captured by the parameter $\Delta$. The special feature of this point in parameter-space is the fact that as the interaction strength $U$ is varied the Fermi-liquid ground state of the system remains invariant. At equilibrium when $U\gtrsim \Gamma$ the system has a pronounced many-body (Abrikosov-Suhl) resonance near the Fermi energy of the lead(s). In the limit $U\rightarrow\infty$ the system maps onto the single-channel Kondo model.

Since the transformation between the LS operators and the electron operators for the interacting problem is nontrivial, we exploit the simple relation in Eq. \eqref{eq:inversion} for the non-interacting case and use this to evaluate the spectral function on the lines of Ref. \onlinecite{DuttLeHur}. We define the {\it effective Hamiltonian} of the system by
\begin{align}
{\cal H}&=\sum_{l=0}^{\infty}\sum_{\alpha=\pm 1}\beta_\alpha \left(\epsilon_k-\frac{\alpha\Phi}{2}\right)\sum_{k=0}^{l}\psi^{\dag(k)}_{\alpha k\sigma}\psi^{(l-k)}_{\alpha k\sigma}.
\label{eq:effective_hamiltonian}
\end{align}
Here, we have expanded the Lippmann-Schwinger states in powers (denoted by the superscript) of the interaction, as given by Eq. (48) in Ref. \onlinecite{DuttLeHur}. The terms corresponding to $l\geq 1$ in Eq.\ \eqref{eq:effective_hamiltonian} encodes the entirety of interaction effects which we denote by ${\cal H}_{\text{int}}$, and define the part ($l=0$) which corresponds to the non-interacting model ${\cal H}^{(0)}$. Furthermore, the inclusion of the inverse temperatures in the definition of the the {\it effective Hamiltonian} circumvents the obstacle in formulating an imaginary-time functional integral due to the absence of a single temperature. This way of defining ${\cal H}$ is effectively a scaling of the energies to make them dimensionless.

The expectation value of an orbitrary operator ${\cal O}$ can be formally expressed as an operator expansion
\begin{widetext}
\begin{align}
\langle{\cal O}\rangle&=\frac{\Tr\left[e^{-{\cal H}}{\cal O}\right]}{\Tr\left[e^{-{\cal H}}\right]}=\sum_{\nu=0}^{\infty}\frac{\Tr\left[e^{-{\cal H}^{(0)}}\left(1+\sum_{m=1}^{\infty}(-1)^m\prod_{i=1}^{m}\left(\int_{0}^{x_{i -1}} dx_{i} e^{x_i {\cal H}^{(0)}} {\cal H}_{\text{int}} e^{-x_i {\cal H}^{(0)}}\right)\right){\cal O}\right]}{\Tr\left[e^{-{\cal H}^{(0)}}\left(1+\sum_{m=1}^{\infty}(-1)^m\prod_{i=1}^{m}\left(\int_{0}^{x_{i -1}} dx_{i} e^{x_i {\cal H}^{(0)}} {\cal H}_{\text{int}} e^{-x_i {\cal H}^{(0)}}\right)\right)\right]}\equiv \sum_{\nu=0}^{\infty}\langle{\cal O}\rangle^{(\nu)}.
\label{eq:formal_expansion}
\end{align}
\end{widetext}
The Green's function for electrons on the dot given by Eq.\ \eqref{eq:retarded} is rewritten  in terms of the non-interacting LS states using the transformations in Eq.\ \eqref{eq:inversion} such that

\begin{align}
G^{\text{ret}}_{d_{\sigma}d^{\dag}_{\sigma}}(\omega)&=\frac{t^2}{\Omega}\sum_{\alpha_{1,2}k_{1,2}}g_{d}(\epsilon_{k_1})g^{\ast}_{d}(\epsilon_{k_2})\notag\\
&\times\left\langle \left\{\frac{1}{\omega+{\cal L}+i\eta}\psi^{(0)}_{\alpha_1 k_1\sigma},\psi^{\dag(0)}_{\alpha_2 k_2\sigma}\right\}\right\rangle.
\label{eq:retarded_GF}
\end{align}
 The simultaneous expansion in powers of ${\cal H}_{\text{int}}$ of the term $(\omega+{\cal L}+i\eta)^{-1}$ and the density matrix, as given by Eq.\ \eqref{eq:formal_expansion}, is carried out systematically below, as in Ref. \onlinecite{DuttLeHur}. 

Using the transformations in Eqns. \ref{eq:inversion} the interaction term can be rewritten as 
\begin{align}
H_{\text{int}}=\frac{U}{2}\left(\frac{t^{2}}{\Omega}\right)^{2}\sum_{1,2,3,4,\sigma}g^{\ast}_{1}g^{\ast}_{2}g_{3}g_{4}\psi^{(0)\dag}_{1\sigma}\psi^{(0)\dag}_{2-\sigma}\psi^{(0)}_{3-\sigma}\psi^{(0)}_{4\sigma},
\label{eq:abb}
\end{align}
where we have introduced the abbreviated notation $l\equiv(\alpha_{l}k_{l})$ and $g_{l}\equiv g_{d}(\ep_{k_{l}})$, implying a sum over the lead index $\alpha$ and the quantum number $k$. Furthermore, we define $\sum_{1}\equiv\sum_{\alpha_{1}k_{1}}$ for the subsequent discussion.

 
In the following discussion it is convenient to absorb the bare energy of the dot in the interaction term, {\it i.e.}, we let $\ep_{d}\rightarrow\tilde{\epsilon}_{d}=\ep_{d}+\frac{U}{2}$, thereby generating an additional term $-\frac{U}{2}\sum_{\sigma}d^{\dag}_{\sigma}d_{\sigma}$. The particle-hole symmetric point is thus specified by $\tilde{\epsilon}_{d}=0$. This redefinition assures that the pole of the bare propagator is correctly positioned, and becomes particularly important when we extend our results to the strong coupling regime. The interaction term is consequently redefined as
\begin{align}
H_{\text{int}}=\frac{U}{2}\left(\widehat{n}_{d}-1\right)^{2}.
\end{align}
Furthermore, we examine deviations from particle-hole symmetry by defining $\Delta=\epsilon_d - U/2$ such that
\begin{align}
g_d(\omega)=\frac{1}{\omega-\Delta+i\Gamma}.
\end{align}
We drop the subscript in $g_d(\omega)$ for notational convenience below. 

Next, we discuss the systematic expansion of the retarded Green's function of the dot in powers of the interaction strength. Recall, 
\begin{align}
G^{\text{ret}}_{d_{\sigma}d^{\dag}_{\sigma}}&(\omega)=\left\langle \left\{\frac{1}{\omega+{\cal L}+i\eta}d_{\sigma},d^{\dag}_{\sigma}\right\}\right\rangle\notag\\
&=\frac{\Tr \left[e^{-\beta(H-Y)}\left\{\frac{1}{\omega+{\cal L}+i\eta}d_{\sigma},d^{\dag}_{\sigma}\right\}\right]}{\Tr\left[e^{-\beta(H-Y)}\right]}.
\end{align}
The exponent is formally expanded according to Eq. \ref{eq:formal_expansion}. Additionally the term $\frac{1}{\omega+{\cal L}+i\eta}d_{\sigma}$ Liouvillian is rewritten as 
\begin{align}
\frac{1}{\omega+{\cal L}+i\eta}d_{\sigma}&=\sum_{k=0}^{\infty}\left(-\frac{1}{\omega+{\cal L}'+i\eta}{\cal L_{\text I}}\right)^{k}\frac{1}{\omega+{\cal L}'+i\eta}d_{\sigma}\notag\\
&\equiv\sum_{k=0}^{\infty}D^{(n)}_{\sigma}(\omega).
\label{eq:expansion_1}
\end{align}

%
The two distinct expansions are then combined to compute the Green's function to the desired order in $H_{\text{int}}$. Note, this expansion is exact in the other parameters such as the tunneling strength $t_{\alpha k}$, the bias voltage $\Phi$ and thermal gradient $\Delta T$. More precisely,
\begin{equation}
G^{\text{ret}}_{d_{\sigma}d^{\dag}_{\sigma}}(\omega)\equiv \sum_{n=0}^{\infty} G^{\text{ret}(n)}_{d_{\sigma}d^{\dag}_{\sigma}}(\omega),
\label{eq:retarded_perturbation_formal}
\end{equation}
where
\begin{equation}
G^{\text{ret}(n)}_{d_{\sigma}d^{\dag}_{\sigma}}(\omega)= \sum_{k=0}^{n} \left\langle \left\{D^{(k)}_{\sigma}(\omega),d^{\dag}_\sigma\right\}\right\rangle^{(n-k)}.
\label{eq:retarded_perturbation}
\end{equation}

In our analysis we will restrict ourselves to ${\cal O}(U^2)$. The 1st-order contribution is explicitly evaluated and the 2nd-order computation is sketched. Further technical details are shown in Appendix \ref{AM}. The non-interacting part follows immediately 
\begin{align}
G^{\text{ret}(0)}_{d^{\dag}_{\sigma}d_{\sigma}}(\omega)&=\left\langle \left\{D^{(0)}_{\sigma},d^{\dag}_{\sigma}\right\}\right\rangle=g(\omega)\left\langle \openone\right\rangle\notag\\
&=\frac{1}{\omega-\Delta+i\Gamma}.
\label{eq:D0}
\end{align}

For the 1st-order parts we have 2 contributions 
\begin{equation}
G^{\text{ret}(1)}_{d_{\sigma}d^{\dag}_{\sigma}}(\omega)= \sum_{k=0}^{1} \left\langle \left\{D^{(k)}_{\sigma}(\omega),d^{\dag}_\sigma\right\}\right\rangle^{(1-k)}.
\label{eq:D1_parts}
\end{equation}
However, since $\left\{D^{(0)}_{\sigma}(\omega),d^{\dag}_\sigma\right\}$ is a c-number its only contribution is to the non-interacting part. Thus, we have
\begin{equation}
G^{\text{ret}(1)}_{d_{\sigma}d^{\dag}_{\sigma}}(\omega)= \left\langle \left\{D^{(1)}_{\sigma}(\omega),d^{\dag}_\sigma\right\}\right\rangle^{(0)}.
\label{eq:D1_parts_b}
\end{equation}

Evaluating the action of the Liouvillians it is straightforward to obtain
\begin{align}
&D^{(1)}_{\sigma}(\omega)
=-U\frac{t}{\sqrt{\Omega}}g(\omega)\bigg(\frac{1}{2}\sum_{1}\frac{g_{1}}{\omega-\ep_{1}+i\eta}\psi^{(0)}_{1\sigma}\bigg)\notag\\
&\qquad-\frac{t^{2}}{\Omega}\sum_{123}\frac{g^{\ast}_{1}g_{2}g_{3}}{\omega+\ep_{1}-\ep_{2}-\ep_{3}+i\eta}\psi^{(0)\dag}_{1-\sigma}\psi^{(0)}_{2-\sigma}\psi^{(0)}_{3\sigma}.
\end{align}
The commutator of interest is given by 
\begin{align}
&\left\{D^{(1)}_{\sigma}(\omega),d^{\dag}_{\sigma}\right\}=Ug(\omega)\bigg[-\frac{1}{2}g(\omega)\notag\\
&\qquad\quad\quad+\frac{t^{2}}{\Omega}\sum_{12}g^{\ast}_{1}g_{2}g(\omega+\ep_{1}-\ep_{2})\psi^{(0)\dag}_{1-\sigma}\psi^{(0)}_{2-\sigma}\bigg].
\end{align}
Using this we determine
\begin{align}
G^{\text{ret}(1)}_{d^{\dag}_{\sigma}d_{\sigma}}(\omega)&=\left\langle \left\{D^{(1)}_{\sigma},d^{\dag}_{\sigma}\right\}\right\rangle^{(0)}\notag\\
&=Ug(\omega)^{2}\bigg[\left(\frac{t^{2}}{\Omega}\right)\sum_{1}\left| g_{1}\right|^{2}f(\ep_{1})-\frac{1}{2}\bigg]\notag\\
&=Ug(\omega)^{2}\bigg[\frac{\Gamma}{\pi}\int d\epsilon\left| g(\epsilon)\right|^{2}f^{\text{eff}}(\ep)-\frac{1}{2}\bigg].
\end{align}
In going from the second to the last line we used the fact that the level spacing goes to zero and the spectrum is linearized around the Fermi energy.

The  ${\cal O}(H_{\text{int}}^2)$ contribution is given by 
\begin{align}
G^{\text{ret}(2)}_{d^{\dag}_{\sigma}d_{\sigma}}(\omega)&=\left\langle \left\{D^{(0)}_{\sigma}(\omega),d^{\dag}_\sigma\right\}\right\rangle^{(2)}+\left\langle \left\{D^{(1)}_{\sigma}(\omega),d^{\dag}_\sigma\right\}\right\rangle^{(1)}\notag\\
&\qquad+\left\langle \left\{D^{(2)}_{\sigma}(\omega),d^{\dag}_\sigma\right\}\right\rangle^{(0)}.
\label{eq:2nd_GF_formal}
\end{align}
Again, the term $\left\langle \left\{D^{(0)}_{\sigma}(\omega),d^{\dag}_\sigma\right\}\right\rangle^{(2)}$ does not contribute. Details of the computation  of the remaining non-trivial terms be found in Appendix \ref{AM}. 

\begin{figure}[t]
	\centering
		\includegraphics[width=0.95\columnwidth]{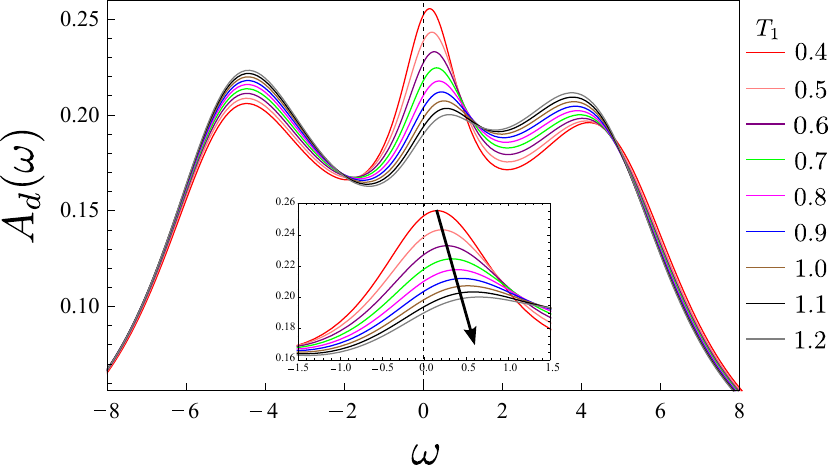}
	\label{fig:molenkamp_spectral_function}
	\caption{Plot of the spectral function of the dot illustrating particle-hole symmetry breaking due to the interplay of $\Delta T$ and $\Phi$. We have used the units $\Gamma=1$ and set $T_{-1}= 0$, $\Phi=0.4$ and $U=3\pi$. Raising the $T_1$ corresponds to increasing both ${\bar T}$ and $\Delta T$, and is seen to suppress the Abrikosov-Suhl resonance as well as amplify the extent of the particle-hole symmetry breaking as evident in the inset.} 
\end{figure} 

\section{Results} 

On increasing the interaction strength $U$, the system eventually enters the strongly correlated regime; the signature of which is the appearance of the many-body Abrikosov-Suhl resonance near the Fermi energy, in addition to the single-particle resonances at energies $\pm U/2$. The behavior of the Abrikosov-Suhl resonance as a function of $\Phi$ has been studied. By applying the Hershfield approach and treating the electron's self-energy to second order in $U$, but treating the bias voltage exactly, our results \cite{DuttLeHur} coincide in the low-bias and high-bias limits to those in similar discussions using the Schwinger-Keldysh scheme \cite{Oguri1,Oguri2,Oguri}. This is akin to the effect of temperature which causes a similar suppression of the many-body resonance \cite{Kaminski,Rosch}. We do not observe a (visible) splitting of the Kondo resonance when increasing the bias voltage. The splitting of the Kondo resonance at intermediate bias voltages is still a subject of debate both theoretically \cite{Anders,Werner,Komnik,Schmidt,Schmidt2,Han} and experimentally \cite{Ensslin}. As already emphasized in the introduction, the interplay of bias voltage and thermal gradient gives rise to a nontrivial effect which we flesh out below; see also Fig. 1.

Due to the Boltzmannian form of the non-interacting density matrix in terms of the non-interacting Lippmann-Schwinger states, which in turn feature in the formal expansion of the spectral function, this gives rise to Fermi functions $f_{\alpha}\left(\epsilon_{k}\right)=(e^{\beta_{\alpha}(\epsilon_{k}-\alpha\Phi/2)}+1)^{-1}$. Furthermore, due to the fact that one always gets a sum over the $\alpha$ indices, the dependence of the spectral function on the thermal gradient and potential bias is dictated exclusively by the symmetric combination of the $f_{\alpha}\left(\epsilon_{k}\right)$'s, which we call the effective Fermi function $f^{\text{eff}}\left(\epsilon_{k}\right)=1/2\sum_{\alpha}f_{\alpha}\left(\epsilon_{k}\right)$. The entirety of the dependence on $\Phi$ and $\Delta T$ in the spectral function (of the electrons on the dot) enter via effective Fermi functions. When $\Delta=0$ and in the absence of biases ({\it i.e.} in equilibrium), particle-hole symmetry implies that $f^{\text{eff}}\left(-\epsilon_{k}\right)=1-f^{\text{eff}}\left(\epsilon_{k}\right)$. 

We now analyze the effect $\Delta T$, $\Phi$ and $\Delta$ on the spectral function of the dot, specifically its influence on the Abrikosov-Suhl resonance. \cite{Kirchner} The case $\Delta\neq 0$ automatically implies the breaking of particle-hole symmetry. However, we find a non-trivial breaking of particle-hole symmetry due to a subtle interplay of the bias voltage and thermal gradients even when $\Delta=0$. For the case of a pure voltage bias  we observe that the particle-hole symmetry is preserved, since under the particle-hole transformation $f^{\text{eff}}\left(\epsilon_{k}\right)$ remains invariant due to the fact that the Fermi functions of the left and right leads are swapped. The case of a pure thermal gradient is considered next, where both $f_{1}\left(\epsilon_{k}\right)$ and $f_{-1}\left(\epsilon_{k}\right)$ are at the same Fermi energy.  In this case it is trivial to see that particle-hole symmetry is preserved. However, under the combined influence of $\Phi$ and $\Delta T$ we observe that this symmetry is broken. One notable consequence of this fact is that $\int d\epsilon_{k} f^{\text{eff}}(\epsilon_{k})\neq1/2$. 

\begin{figure}[t]
	\centering
		\includegraphics[width=0.95\columnwidth]{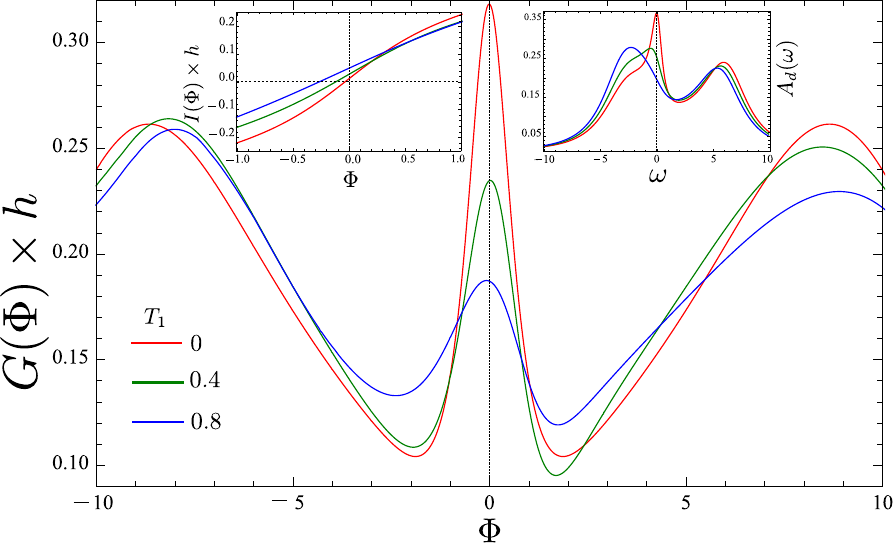}
	\label{fig:conductivity}
	\caption{Effect of $\Delta T$ on the differential conductance when $\Delta=0.4$. We have used the units $\Gamma=1$ and taken $U=3\pi$. The inset on right shows the corresponding spectral function when $\Phi=0$ and the current in the neighborhood of zero-bias is shown in the left inset.} 
\end{figure}

In order to estimate the extent of this feature, in particular its effect on the Abrikosov-Suhl resonance, we compute the self-energy of the electrons on the dot to second-order in the interaction strength and exact in all other parameters. To do so we use the Dyson equation 
\begin{equation}
G^{\text{ret}}_{d_{\sigma}d^{\dag}_{\sigma}}(\omega)=G^{\text{ret}(0)}_{d_{\sigma}d^{\dag}_{\sigma}}(\omega)+G^{\text{ret}(0)}_{d_{\sigma}d^{\dag}_{\sigma}}(\omega)\Sigma(\omega)G^{\text{ret}(0)}_{d_{\sigma}d^{\dag}_{\sigma}}(\omega),
\label{eq:Dyson}
\end{equation}
in conjunction with the perturbative expansion of the Green's function in Eq.\ \eqref{eq:retarded_perturbation_formal}. This allows us to compute the $n$-th order ($n\geq 1$) contribution in U to the self energy of the system via the relation
\begin{align}
\Sigma^{(n)}=\frac{1}{g(\omega)^2}G^{\text{ret}(n)}_{d_{\sigma}d^{\dag}_{\sigma}}(\omega)
\label{eq:self_energy_pert}
\end{align}
Here $\Sigma(\omega)$ denotes the {\bf complete} self energy of the system. However for the Born Approximation we need to evaluate the {\bf proper} self energy of the system $\Sigma^{\star}$ which satisfies
\begin{equation}
G^{\text{ret}}_{d_{\sigma}d^{\dag}_{\sigma}}(\omega)=G^{\text{ret}(0)}_{d_{\sigma}d^{\dag}_{\sigma}}(\omega)+G^{\text{ret}(0)}_{d_{\sigma}d^{\dag}_{\sigma}}(\omega)\Sigma^{\star}(\omega)G^{\text{ret}}_{d_{\sigma}d^{\dag}_{\sigma}}(\omega).
\label{eq:Dyson_proper}
\end{equation}
In order to extract this contribution we must discard the terms corresponding to irreducible Feynman diagrams in the Green's function expansion of Eq.\ \eqref{eq:retarded_perturbation_formal}. Details are given in Appendix \ref{Appendix_final}.

The deviation from particle-hole symmetry emerges at the first order itself
\begin{equation}
\Sigma^{\star(1)}=U\bigg[\frac{\Gamma}{\pi}\int d\epsilon\left| g_d(\epsilon)\right|^{2}f^{\text{eff}}(\epsilon)-\frac{1}{2}\bigg],
\label{eq:self_1st_proper}
\end{equation}
which is in general non-zero even if $\epsilon_d=-U/2$. The strongly correlated dynamics can only be captured by terms proportional to ${\cal O}(U^2)$ and higher. We compute the proper self-energy to second order to obtain

\begin{widetext}
\begin{align}
&\Sigma^{\star(2)}=U^2\bigg[\frac{\Gamma}{\pi}\int d\epsilon \frac{ f^{\text{eff}}(\epsilon)\left|g_d(\epsilon)\right|^2}{\omega+\epsilon-2\Delta+2i\Gamma}+\left(\frac{\Gamma}{\pi}\right)^2\int d\epsilon_1 d\epsilon_2 \left|g_d(\epsilon_1)\right|^2\left|g_d(\epsilon_2)\right|^2 f_{1}^{\text{eff}}f_{2}^{\text{eff}}\bigg\{g_d(\omega-\epsilon_1-\epsilon_2)-2g_d(\omega+\epsilon_1-\epsilon_2)\bigg\}\notag\\
&\qquad\qquad\qquad\qquad+\frac{\Sigma^{\star(1)}}{U}\left\{\frac{\Gamma}{\pi}\int d\epsilon f^{\text{eff}}(\epsilon)\left(\frac{2\left(\epsilon-\Delta\right)}{(\left(\epsilon-\Delta\right)^2+\Gamma^2)^2}-\left(\frac{\left|g_d(\epsilon)\right|^2}{\omega-\epsilon-2\Delta+2i\Gamma}-\frac{\left|g_d(\epsilon)\right|^2}{\omega-\epsilon+2i\Gamma}\right)\right)\right\}\bigg].
\end{align}
\end{widetext}
It is straightforward to observe that on setting $\Delta T=0$ we obtain the out-of-equilibrium result of Ref. \onlinecite{DuttLeHur}, where only potential biases were considered.

Next, we analyze the system using the Born-approximation for the Green's fuction in Eq. \eqref{eq:retarded}, by applying an extension of the method in Ref. \onlinecite{DuttLeHur}. In this case we incorporate two different temperatures of the bath via Eq. \eqref{eq:formal_expansion}. The spectral function of the dot is shown in Fig. 1, where a deviation from particle-hole symmetry is apparent. We set the unit of energy by $\Gamma$ and consider the case when the interaction strength $U=3\pi$. Additionally, we fix $\Phi=0.4$ and $T_{-1}=0.0$, and vary the temperature of the left lead $T_1$. The mean temperature $\bar T$ determines the height of the Abrikosov-Suhl resonance which is inhibited as $\bar T$ is increased. Another interesting feature is the migration of the Abrikosov-Suhl peak as $\bar T$ is changed, an effect which can be explained on the basis of the underlying Fermi Liquid expansion of the spectral function, where terms coupling $\Phi$ and $\Delta T$ grow as $\bar{T}$ increases. An experiment to map the spectral function of the dot has been performed in Ref. \onlinecite{Ensslin}. The combined effect of $\Phi$ and $\Delta T$ we analyze can be probed using this setup. 

Using the expression for the current in Eq.\ \eqref{eq:meir_wingreen} we compute the differential (electrical) conductivity $G(\Phi)= e^{-1} \partial_{\Phi} I(\Phi, \Delta T)$, where we evaluate the current using the Born approximation. In Fig. 2 we set $\Delta=0.4$, thus breaking particle-hole symmetry explicitly. We then study the effect of a temperature gradient on $G(\Phi)$. Here, we note that this causes a migration of the Abrikosov-Suhl peak as evident in the inset showing the spectral function. This effect is manifested in the current which now has a non-zero contribution at zero-bias driven by thermal effects. The Fermi functions in the expression for current in Eq.\ \eqref{eq:meir_wingreen} enhances this effect. This analysis qualitatively captures the results of Ref. \onlinecite{Molenkamp}. Further technical details are provided in Appendix \ref{AM}.

\section{Conclusion}

To summarize, we have developed a formalism to describe thermal gradients as well as bias voltage effects in quantum impurity models beyond the linear
response regime. The scope of this formalism can be expanded and used to focus on other aspects of the system, for example, the thermopower and the thermal conductivity.  Furthermore, we have predicted that the interplay between thermal gradients and bias voltages gives rise to an additional particle-hole symmetry breaking in the profile of the localized electron's spectral function, a feature which can be directly probed experimentally. Note, this feature is to be contrasted with particle-hole symmetry breaking commonly encountered in literature which involves a deviation of $\epsilon_d$, the single particle energy of the dot electron, from the value $-U/2$.  

\section{Acknowledgements}
We acknowledge discussions with M. B\"uttiker, N. Andrei, C. Aron, A. Georges, C. Grenier, J. E. Han, J. Koch, L. Messio, P. Orth, O. Parcollet, H. Tureci and Z. Ristivojevic.
This work was supported by the Department of Energy under the grant DE-FG02-08ER46541 and by the Labex PALM at Paris-Saclay.

\appendix

\section{Derivation of Thermoelectric Steady-State Density Matrix}\label{derivation_dm}
We consider a system consisting of two Fermi-liquid leads, coupled by tunnel junctions to a central system with a number of discrete levels (QD). The Hamiltonian of this generic system can be written in the form $\hat{H}=\hat{H}_{L}+\hat{H}_{D}+\hat{H}_{T}$. The first term
\begin{equation}
\hat{H}_{L}=\sum_{\alpha k \sigma}\epsilon_{\alpha k}c^{\dag}_{\alpha k\sigma}c_{\alpha k\sigma}=\hat{H}_{L_1}+\hat{H}_{L_{-1}}
\label{eq:lead_hamiltonian}
\end{equation}
describes the left and right leads ($\alpha=\pm1$), where $c_{\alpha k\sigma}$ ($c_{\alpha k\sigma}^\dag)$ annihilates (creates) an electron (strictly speaking a Fermi-liquid quasiparticle) in state $k$ with spin projection $\sigma$ in lead $\alpha$. The corresponding energy dispersion is denoted by $\epsilon_{\alpha k}$. These leads couple to the dot, whose Hamiltonian is given by
\begin{equation}
\hat{H}_{D}=\ \sum_{\sigma}\epsilon_{d}d^{\dag}_{\sigma}d_{\sigma}+\hat{H}_\text{int},
\label{eq:dot_hamiltonian}
\end{equation}
with $d_{\sigma}$ ($d_{\sigma}^\dag$) annihilating (creating) an electron with spin $\sigma$ in the  discrete level with energy $\epsilon_d$. Any electron interactions are lumped into the contribution $H_\text{int}$, and we will assume that these interactions are localized on the dot (as it is the case for the Anderson model). Finally, the tunneling of electrons between leads and dot is captured by the tunneling Hamiltonian
\begin{equation}
\hat{H}_{T}=\frac{1}{\sqrt{\Omega}}\sum_{\alpha k \sigma}t_{\alpha k}\left(c^{\dag}_{\alpha k\sigma}d_{\sigma}+\text{h.c.}\right),
\label{eq:tunneling}
\end{equation}
where $\Omega$ is the lead volume (assumed identical for both leads) and $t_{\alpha k}$ specifies the tunneling matrix element for electron transfer between state $k$ in lead $\alpha$ and the discrete dot state.

We assume at an initial time $t\rightarrow -\infty$, the tunneling between the leads and dot is absent, and the left and right leads are at an inverse temperature $\beta_1$ and $\beta_{-1}$ respectively. Assume, without loss of generality, that $\beta_1<\beta_{-1}$ ({\it i.e.} $T_1>T_{-1}$). We introduce the bias $\Phi$ and write $\mu_1=\Phi/2$ and $\mu_{-1}=-\Phi/2$.  The initial density matrix
\begin{equation}
\rho_0=\exp\left[-\left(\beta_1 \hat{H}_{L_1}+\beta_{-1} \hat{H}_{L_{-1}}\right)\right]\otimes\rho_0^{D}.
\label{eq:initial_dm_te}
\end{equation}
Defining $\beta_0=\left(\beta_{1}+\beta_{-1}\right)/2$ and $\tilde{\Delta}=\left(\beta_{-1}-\beta_1\right)$ this can be rewritten as
\begin{align}
\rho_0=\exp\left[-\beta_0\left(\hat{H}_0-\hat{W}_0\right)\right],
\end{align}
where the operator
\begin{equation}
{\hat H}_0=\hat{H}_L+\hat{H}_D,
\label{eq:without_tunn}
\end{equation}
and
\begin{align}
\hat{W}_0=\frac{1}{2}\sum_{\alpha k\sigma}\left[\alpha\frac{\tilde{\Delta}}{\beta_0}\epsilon_{\alpha k}+\Phi\left(\alpha-\frac{\tilde{\Delta}}{2\beta_0}\right)\right]c^{\dag}_{\alpha k\sigma}c_{\alpha k\sigma}.
\end{align}
Formally this is identical to the case when only a bias voltage is present (with modified single-particle energies), which has been analyzed in detail in Appendix A of Ref. \onlinecite{DuttLeHur}. The tunneling($\hat{H}_T$) is introduced adiabatically such that we obey the open-system limit steady state which  in turn guarantees the formation of the steady-state. The steady state density matrix is therefore
\begin{equation}
\hat{\rho}=\exp\left[-\beta_0\left(\hat{H}-\hat{W}\right)\right],
\label{eq:density_matrix_te}
\end{equation}
where
\begin{equation}
\hat{W}=\frac{1}{2}\sum_{\alpha k\sigma}\left[\alpha\frac{\tilde{\Delta}}{\beta_0}\epsilon_{\alpha k}+\Phi\left(\alpha-\frac{\tilde{\Delta}}{2\beta_0}\right)\right] \psi^{\dag}_{\alpha k\sigma}\psi_{\alpha k\sigma}.
\label{eq:w_operator}
\end{equation}
We can rewrite Eq.\ \eqref{eq:density_matrix_te} as 
\begin{align}
\rho=\exp\left[-\beta_0\sum_{\alpha k\sigma}\tilde{E}_{\alpha k}\psi^\dagger_{\alpha k\sigma}\psi_{\alpha k\sigma}\right],
\end{align}
where the effective energy spectrum has additional lead dependence 
\begin{align}
\tilde{E}_{\alpha k}=\left(1-\alpha\frac{\tilde{\Delta}}{2\beta_0}\right)\epsilon_{\alpha k}-\frac{\Phi}{2}\left(\alpha-\frac{\tilde{\Delta}}{2\beta_0}\right).
\end{align} 
On simplification of the expression for the density matrix we find
\begin{align}
\rho&=\exp\left[-\sum_{\alpha k\sigma}\beta_{\alpha}\left(\epsilon_{\alpha k}-\alpha\frac{\Phi}{2}\right)\psi^\dagger_{\alpha k\sigma}\psi_{\alpha k\sigma}\right]\notag\\
&=\prod_{\alpha=\pm 1}\otimes\exp\left[-\beta_\alpha\sum_{k\sigma}\left(\epsilon_{\alpha k}-\frac{\alpha\Phi}{2}\right)\psi^\dagger_{\alpha k\sigma}\psi_{\alpha k\sigma}\right].
\label{eq:electro_thermal_dm}
\end{align}
We will drop the circumflex on the operators in the ensuing sections.

\section{Local description of transport}\label{observables}
We consider the current flowing through the dot, a transport observable of prime interest. It is given by
\begin{align}
I&=\frac{I_{1}-I_{-1}}{2}=-\frac{e}{2}\left\langle\frac{d\left(N_{1}(t)-N_{-1}(t)\right)}{dt}\right\rangle\notag\\
&=i\frac{e}{2\hbar}\sum_{\alpha}\alpha\left\langle[N_{\alpha}(t),H]\right\rangle\notag\\
&=i\sum_{\alpha k\sigma}\alpha \frac{et_{\alpha k}}{2\hbar\sqrt{\Omega}}\left\langle\left(c^{\dag}_{\alpha k\sigma}d_{\sigma}-d^{\dag}_{\sigma}c_{\alpha k\sigma}\right)\right\rangle\notag\\
&=\frac{e}{\hbar}\sum_{\alpha k \sigma}\alpha\frac{t_{\alpha k}}{\sqrt{\Omega}}\iim\left[\mathcal{ G}_{c_{\alpha k \sigma}d_{\sigma}^\dag}(\tau=0)\right],
\label{current_primary}
\end{align}
where $e$ denotes the charge of the electron. Similarly for the energy-current in steady state($E_{\pm 1}$ denote the total energy of the $\alpha=\pm 1$ leads respectively) ,
\begin{align}
I_E&=\frac{I^E_{1}-I^E_{-1}}{2}=-\frac{1}{2}\left\langle\frac{d\left(E_{1}(t)-E_{-1}(t)\right)}{dt}\right\rangle\notag\\
&=\frac{i}{2\hbar}\sum_{\alpha}\alpha\left\langle[E_{\alpha}(t),H]\right\rangle\notag\\
&=i\sum_{\alpha k\sigma}\alpha \frac{t_{\alpha k}}{2\hbar\sqrt{\Omega}}\epsilon_{\alpha k}\left\langle\left(c^{\dag}_{\alpha k\sigma}d_{\sigma}-d^{\dag}_{\sigma}c_{\alpha k\sigma}\right)\right\rangle\notag\\
&=\frac{1}{\hbar}\sum_{\alpha k \sigma}\alpha\epsilon_{\alpha k}\frac{t_{\alpha k}}{\sqrt{\Omega}}\iim\left[\mathcal{ G}_{c_{\alpha k \sigma}d_{\sigma}^\dag}(\tau=0)\right].
\label{heat_current_primary}
\end{align}
Here we define the {\bf modified imaginary-time evolution} of an operator
\begin{align}
{\cal O}(\tau)=U^{\dag}(\tau){\cal O}(0)U(\tau),
\label{eq:imaginary_time_evolution}
\end{align}
where we have defined the time evolution operator in imaginary-time
\begin{equation}
U(\tau)=\left(\prod_{\alpha=\pm 1}\otimes\exp\left[-\tau\sum_{k\sigma}\left(\epsilon_{\alpha k}-\frac{\alpha\Phi}{2}\right)\psi^\dagger_{\alpha k\sigma}\psi_{\alpha k\sigma}\right]\right)
\label{eq:time_evolution}
\end{equation}
and subsequently the {\bf modified imaginary-time Green's function}
\begin{align}
\mathcal{ G}_{{\cal O}_1 {\cal O}_2}(\tau)=-\left\langle{\cal T}_{\tau}\left[{\cal O}_1(\tau) {\cal O}_2(0)\right]\right\rangle.
\end{align} 
Note, all expectation values are taken with respect to the steady-state effective equilibrium density matrix in Eq.\ \eqref{eq:electro_thermal_dm}.

Here, we consider the general setup, where the electrons on the dot are interacting. For Coulomb interactions we have the familiar single-impurity Anderson model. In this Section we reformulate the transport quantities in terms of the local spectral function of the quantum impurity and obtain Meir-Wingreen like formulae in the more general context of leads with different temperatures.

From Eq.\ \eqref{current_primary} we obtain
\begin{align}
I&=\frac{e}{\hbar}\sum_{\alpha k \sigma}\alpha\frac{t_{\alpha k}}{\sqrt{\Omega}}\iim\left[\mathcal{ G}_{c_{\alpha k \sigma}d_{\sigma}^\dag}(\tau=0^{+})\right]\notag\\
&=-\frac{e}{\hbar}\sum_{\alpha k \sigma}\alpha\frac{t_{\alpha k}}{\sqrt{\Omega}}\iim\left[\left\langle c_{\alpha k \sigma}(\tau=0^{+})d_{\sigma}^\dag\right\rangle \right].
\end{align}
Let us focus on the term
\begin{align}
&\left\langle c_{\alpha k \sigma}(\tau=0^{+})d_{\sigma}^\dag\right\rangle= \left\langle \psi_{\alpha k \sigma}(\tau=0^{+})d_{\sigma}^\dag\right\rangle\notag\\
&\qquad-\frac{t_{\alpha k}}{\sqrt{\Omega}}\left\langle \frac{1}{\epsilon_{\alpha k} +{\cal L}-{\cal L}_{Y}}d_{\sigma}(\tau=0^{+})d_{\sigma}^\dag\right\rangle. 
\end{align}
Assume that tunneling is energy-independent, {\it i.e.} $t_{\alpha k}=t_{\alpha}$ (we have already assumed that the leads are identical, {\it i.e.} $\epsilon_{\alpha k}=\epsilon_k$). In steady-state we see $I_1=I_{-1}$ so we can define $I=\left(t_{-1}^2 I_1 +t_{-1}^2 I_{-1}\right)/\left(t_{1}^2+t_{-1}^2\right)$ which renders the second part irrelevant. We define the actions of the Liouvillians ${\cal L}^{\alpha}$ and ${\cal L}^{\alpha}_{Y}$ on an operator ${\cal O}$ by ${\cal L}^{\alpha}{\cal O}=\left[\sum_{k\sigma}\epsilon_{k\sigma}\psi^{\dag}_{\alpha k\sigma}\psi_{\alpha k\sigma},{\cal O}\right]$ and ${\cal L}^{\alpha}_{Y}{\cal O}=\left[\sum_{k\sigma}\alpha\frac{\Phi}{2}\psi^{\dag}_{\alpha k\sigma}\psi_{\alpha k\sigma},{\cal O}\right]$. The Fourier transform (the temperature index of which is unambiguous for the $\psi_{\alpha k\sigma}$) is given by
\begin{align}
\mathcal{ G}_{\psi_{\alpha k \sigma}d_{\sigma}^\dag}(i\omega_n^{\alpha})&=\int_{0}^{\beta_{\alpha}}d\tau e^{i\omega_n^{\alpha}\tau}\mathcal{ G}_{\psi_{\alpha k \sigma}d_{\sigma}^\dag}(\tau)\notag\\
&=-\int_{0}^{\beta_{\alpha}}d\tau \left\langle e^{i\omega_n^{\alpha}\tau}e^{({\cal L}^{\alpha}-{\cal L}^{\alpha}_Y)\tau}\psi_{\alpha k \sigma}d^\dag_{\sigma}\right\rangle\notag\\
&=-\left\langle \frac{e^{(i\omega_n^{\alpha}+{\cal L}^{\alpha}-{\cal L}^{\alpha}_Y)\tau}}{(i\omega_n^{\alpha}+{\cal L}-{\cal L}_Y)}\bigg|_{0}^{\beta_{\alpha}}\psi_{\alpha k \sigma}d^\dag_{\sigma}\right\rangle\notag\\
&=\left\langle \frac{e^{({\cal L}^{\alpha}-{\cal L}^{\alpha}_Y)\beta_{\alpha}}}{i\omega_n^{\alpha}+{\cal L}^{\alpha}-{\cal L}^{\alpha}_Y}\psi_{\alpha k \sigma}d^\dag_{\sigma}\right\rangle\notag\\
&\qquad+\left\langle \frac{1}{i\omega_n^{\alpha}+{\cal L}^{\alpha}-{\cal L}^{\alpha}_Y}\psi_{\alpha k \sigma}d^\dag_{\sigma}\right\rangle\notag\\
&=\left\langle \frac{e^{(H^{\alpha}-Y^{\alpha})\beta_{\alpha}}}{i\omega_n^{\alpha}+{\cal L}^{\alpha}-{\cal L}^{\alpha}_Y}\psi_{\alpha k \sigma}e^{-(H^{\alpha}-Y^{\alpha})\beta_{\alpha}}d^\dag_{\sigma}\right\rangle\notag\\
&\qquad+\left\langle \frac{1}{i\omega_n^{\alpha}+{\cal L}^{\alpha}-{\cal L}^{\alpha}_Y}\psi_{\alpha k \sigma}d^\dag_{\sigma}\right\rangle\notag\\
&=\left\langle \left\{\frac{1}{i\omega_n^{\alpha}+{\cal L}^{\alpha}-{\cal L}^{\alpha}_Y}\psi_{\alpha k \sigma},d^\dag_{\sigma}\right\}\right\rangle.
\end{align}
In the last step we used the cyclic property of the trace to obtain the anticommutator. Thus we get
\begin{align}
\mathcal{ G}_{\psi_{\alpha k\sigma}d_{\sigma}^\dag}(i\omega_n^{\alpha})=\frac{1}{i\omega_n^{\alpha}-\epsilon_k+\alpha\frac{\Phi}{2}}\left\langle \left\{\psi_{\alpha k \sigma},d^\dag_{\sigma}\right\}\right\rangle,
\end{align}
and subsequently
\begin{align}
\mathcal{ G}_{\psi_{\alpha k\sigma}d_{\sigma}^\dag}(0^{+})&=\frac{1}{\beta_{\alpha}}\sum_{\omega_n^{\alpha}}\frac{e^{i\omega_n^{\alpha}0^{+}}}{i\omega_n^{\alpha}-\epsilon_k+\alpha\frac{\Phi}{2}}\left\langle \left\{\psi_{\alpha k \sigma},d^\dag_{\sigma}\right\}\right\rangle\notag\\
=\frac{t_{\alpha}}{\sqrt{\Omega}}&f_{\alpha}(\epsilon_k)\left\langle \left\{d_{\sigma},\frac{1}{\epsilon_k - {\cal L} +i\eta}d^\dag_{\sigma}\right\}\right\rangle.
\end{align}
The current can be thus written as
\begin{align}
I&=\frac{2e}{\hbar}\frac{\Gamma_1 \Gamma_{-1}}{\Gamma_1+\Gamma_{-1}}\sum_{\alpha \sigma} \alpha\frac{1}{\pi}\int d\epsilon_k f_{\alpha}(\epsilon_k)\notag\\&\qquad\qquad\times\iim \left[\left\langle \left\{d_{\sigma},\frac{1}{\epsilon_k - {\cal L} +i\eta}d^\dag_{\sigma}\right\}\right\rangle\right].
\label{eq:a}
\end{align}
The real-time Green's function
\begin{align}
G^{\text{ret}}_{{\cal O}_1{\cal O}_2}(t)=-i\theta(t)\left\langle \left\{{\cal O}_1(t){\cal O}_2\right\}\right\rangle,
\end{align}
where the operator ${\cal O}_1(t)=e^{i H t}{\cal O}_1e^{-i H t}$ is in the Heisenberg picture. The Fourier transform is given by
\begin{align}
G^{\text{ret}}_{{\cal O}_1{\cal O}_2}(\omega)&=\int_{0}^{\infty}dt e^{i(\omega+i\eta)t} G^{\text{ret}}_{{\cal O}_1{\cal O}_2}(\omega)\notag\\
&=\left\langle \left\{{\cal O}_1,\frac{1}{\omega - {\cal L} +i\eta}{\cal O}_2\right\}\right\rangle.
\end{align}
Comparing with Eq.\ \eqref{eq:a} we obtain
\begin{align}
\label{eq:current_Meir_Wingreen}
I&=\frac{2 e}{\hbar}\frac{\Gamma_1 \Gamma_{-1}}{\Gamma_1+\Gamma_{-1}}\sum_{\alpha \sigma}\alpha \int d\epsilon_k f_{\alpha}(\epsilon_k)A_d(\epsilon_k),
\end{align}
where we have defined the spectral function
\begin{align}
A_d(\omega)=-\frac{1}{\pi}\iim \left[G^{\text{ret}}_{d_{\sigma}d^{\dag}_{\sigma}}(\omega)\right].
\end{align}
In general, the spectral function is an explicit function of the bias $\Phi$ and the two temperatures ($T_1$ and $T_{-1}$) of the leads. By identical reasoning for the energy-current we get
\begin{align}
\label{eq:heat_current_Meir_Wingreen}
I_E=\frac{2 e}{\hbar}\frac{\Gamma_1 \Gamma_{-1}}{\Gamma_1+\Gamma_{-1}}\sum_{\alpha \sigma}\alpha \int d\epsilon_k \epsilon_k f_{\alpha}(\epsilon_k)A_d(\epsilon_k).
\end{align}
Another, quantity of interest is the thermopower $S\equiv e^{-1}\left(\Phi/\Delta T\right)|_{I=0}$. To derive the thermopower we expand Eq. \eqref{eq:current_Meir_Wingreen} to linear order in $\Delta T$ and $\Phi$ (assume identical leads for the moment)
\begin{align} 
&I=\frac{e\Gamma}{2\hbar}\sum_{\alpha \sigma}\alpha \int d\epsilon_k \bigg\{f(\epsilon_k)\bigg[\partial_\Phi A_d(\epsilon_k)\big|_{\Phi=0,\Delta T=0}\Phi\notag\\
&+\partial_{\Delta T} A_d(\epsilon_k)\big|_{\Phi=0,\Delta T=0}\Phi\bigg]+A_d(\epsilon_k)\big|_{\Phi=0,\Delta T=0}\times\notag\\
&\left[\partial_\Phi f_\alpha(\epsilon_k)\big|_{\Phi=0,\Delta T=0}\Phi+\partial_{\Delta T} f_\alpha(\epsilon_k)\big|_{\Phi=0,\Delta T=0}\Delta T\right]\bigg\}. 
\end{align}
Note that $A_d(\epsilon_k)\big|_{\Phi=0,\Delta T=0}$ is the spectral function of the electrons on the QD at equilibrium, which we will refer to as $A^{\text{eq}}_d(\epsilon_k)$. The first term is zero, and the second term can be simplified using the relations $\partial_\Phi f_\alpha(\epsilon_k)\big|_{\Phi=0,\Delta T=0}=-\frac{\alpha}{2}\partial_{\epsilon_k} f(\epsilon_k)$ and $\partial_\Phi f_\alpha(\epsilon_k)\big|_{\Phi=0,\Delta T=0}=-\beta\epsilon_k\partial_{\epsilon_k}f(\epsilon_k)$ we obtain
\begin{align} 
I&=-\frac{e\Gamma}{2\hbar}\sum_{\sigma}\int d\epsilon_k A^{\text{eq}}_d(\epsilon_k)\left[\partial_{\epsilon_k} f(\epsilon_k)\Phi+\beta\epsilon_k\partial_{\epsilon_k}f(\epsilon_k)\Delta T\right]. 
\end{align}

\section{Thermoelectric transport through the Anderson Impurity}
\label{AM}

In this Section, we investigate thermoelectric transport through the Anderson impurity model - a commonly encountered experimental scenario and serves as the microscopic model for a wide variety of fundamental physical phenomena. The interaction is given by
\begin{align}
H_{\text{int}}=\frac{U}{2}\widehat{n}_{d}\left(\widehat{n}_{d}-1\right)=\frac{U}{2}\sum_{\sigma}d^{\dag}_{\sigma}d^{\dag}_{-\sigma}d_{-\sigma}d_{\sigma},
\end{align} 
and denotes the Coulomb interaction between electrons of opposite spin projections on the quantum dot.  Here, $\widehat{n}_{d}=\sum_{\sigma}d^{\dag}_{\sigma}d_{\sigma}$ denotes the electron number operator of the dot. 

As a starting point we consider the case when $U=0$.  Here, it is simple to relate the scattering state operators with those corresponding with degrees of freedom of the system
\begin{align}
d_{\sigma}^{\dag}=\frac{t}{\sqrt{\Omega}}\sum_{\alpha k}g^{\ast}_{d}(\epsilon_{k})\psi_{\alpha k\sigma}^{(0)\dag}.
\label{eq:d expansion}
\end{align}
and
\begin{align}
c^\dagger_{\alpha k\sigma}=\psi^{(0)\dagger}_{\alpha k\sigma}-
\frac{t^{2}}{\Omega}\sum_{\alpha'k'}\frac{g^{\ast}_{d}(\epsilon_{k'})}{\epsilon_{k}-\epsilon_{k'}+i\eta}\psi_{\alpha'k'\sigma}^{(0)\dag}
\label{eq:c expansion},
\end{align}
where we have defined 
\begin{align}
g_{d}(\epsilon_{k})=\frac{1}{\epsilon_{k}-\epsilon_{d}+i\Gamma}.
\label{eq:gd}
\end{align}
In the above steps we have assumed that the tunneling amplitude is energy-independent and is the same for each channel, {\it i.e.} $t_{\alpha\beta}=t$. Furthermore, we have assumed that the energy spectrum for individual channels is identical, , {\it i.e.} $\epsilon_{\alpha k}=\epsilon_{k}$.

Thus, the interaction can be rewritten as 
\begin{align}
H_{\text{int}}=\frac{U}{2}\left(\frac{t^{2}}{\Omega}\right)^{2}\sum_{1,2,3,4,\sigma}g^{\ast}_{1}g^{\ast}_{2}g_{3}g_{4}\psi^{(0)\dag}_{1\sigma}\psi^{(0)\dag}_{2-\sigma}\psi^{(0)}_{3-\sigma}\psi^{(0)}_{4\sigma},
\label{eq:abb}
\end{align}
where we have introduced the abbreviated notation $l\equiv(\alpha_{l}k_{l})$ and $g_{l}\equiv g_{d}(\ep_{k_{l}})$, implying a sum over the lead index $\alpha$ and the quantum number $k$. Furthermore, we define $\sum_{1}\equiv\sum_{\alpha_{1}k_{1}}$ for the subsequent discussion.

We study quantum transport through an Anderson impurity in the vicinity of the particle-hole symmetric point, which is given by $\ep_{d}=-\frac{U}{2}$. The special feature of the particle-hole symmetric point is that by tuning the interaction strength $U$, one can pass smoothly from the weak to the strong coupling regime such that the Fermi liquid fixed point of the Hamiltonian remains invariant. Also, the Anderson impurity model maps exactly onto the Kondo model in the limit $U\rightarrow\infty$ at this point in parameter space. 
 
In the following discussion it is convenient to absorb the bare energy of the dot in the interaction term, {\it i.e.}, we let $\ep_{d}\rightarrow\tilde{\epsilon}_{d}=\ep_{d}+\frac{U}{2}$, thereby generating an additional term $-\frac{U}{2}\sum_{\sigma}d^{\dag}_{\sigma}d_{\sigma}$. The particle-hole symmetric point is thus specified by $\tilde{\epsilon}_{d}=0$. This redefinition assures that the pole of the bare propagator is correctly positioned, and becomes particularly important when we extend our results to the strong coupling regime. The interaction term is consequently redefined as
\begin{align}
H_{\text{int}}=\frac{U}{2}\left(\widehat{n}_{d}-1\right)^{2}.
\end{align}
Furthermore, we examine deviations from particle-hole symmetry by defining $\Delta=\epsilon_d - U/2$ such that
\begin{align}
g_d(\omega)=\frac{1}{\omega-\Delta+i\Gamma}.
\end{align}
We drop the subscript in $g_d(\omega)$ for notational convenience below.

\begin{widetext}
\subsection{Perturbative computation of the retarded QD Green's function}
Below, we outline the perturbative expansion of the retarded Green's function of electrons on the dot in powers of the interaction
\begin{align}
G^{\text{ret}}_{d_{\sigma}d^{\dag}_{\sigma}}&(\omega)=\left\langle \left\{\frac{1}{\omega+{\cal L}+i\eta}d_{\sigma},d^{\dag}_{\sigma}\right\}\right\rangle=\frac{\Tr \left[e^{-\beta(H-Y)}\left\{\frac{1}{\omega+{\cal L}+i\eta}d_{\sigma},d^{\dag}_{\sigma}\right\}\right]}{\Tr\left[e^{-\beta(H-Y)}\right]}.
\end{align}
We systematically collect powers of the interaction arising from both the expansion of the exponent containing the action and also the Liouvillian. Let us first compute $\frac{1}{\omega+{\cal L}+i\eta}d_{\sigma}$.
\begin{align}
\frac{1}{\omega+{\cal L}+i\eta}d_{\sigma}=\sum_{k=0}^{\infty}\left(-\frac{1}{\omega+{\cal L}'+i\eta}{\cal L_{\text I}}\right)^{k}\frac{1}{\omega+{\cal L}'+i\eta}d_{\sigma}\equiv\sum_{k=0}^{\infty}D^{(n)}_{\sigma}(\omega).
\label{eq:expansion_1}
\end{align}
This can be written recursively as
\begin{align}
D^{(k)}_{\sigma}(\omega)=-\frac{1}{\omega+{\cal L}'+i\eta}{\cal L_{I}}D^{(k-1)}_{\sigma}(\omega).
\label{eq:recursion}
\end{align}
Here we have defined the Liouvillians ${\cal L}'$ and ${\cal L_{I}}$ such that it's action on an operator ${\cal O}$ is given by ${\cal L}'{\cal O}=[H-H_{\text{int}},{\cal O}]$ and ${\cal L}_{I}{\cal O}=[H_{\text{int}},{\cal O}]$. 

Let us now separate the {\it effective Hamiltonian} into (a)the non-interacting part and (b)the interaction dependent part
\begin{align}
{\cal H}&=\sum_{\alpha=\pm 1}\beta_\alpha (H_\alpha-Y_\alpha)=\sum_{\alpha=\pm 1}\beta_\alpha \left(\epsilon_k-\frac{\alpha\Phi}{2}\right)\sum_{k\sigma}\psi^{\dag}_{\alpha k\sigma}\psi_{\alpha k\sigma}
\label{eq:eff_inter}
\end{align}
We expand the Lippmann-Schwinger operators of the interacting system in terms of the Lippmann-Schwinger operators of the noninteracting model
\begin{align}
\psi^{\dag}_{\alpha k\sigma}=\psi^{\dag(0)}_{\alpha k\sigma}+\frac{t^2}{\Omega}\sum_{l=1}^{\infty}\left[\frac{1}{\epsilon_k-{\cal L}'+i\eta}{\cal L}_I\right]^l\sum_{\alpha'k'}\frac{g^{\ast}_d(\epsilon_k)}{\epsilon_k-\epsilon_k'+i\eta}\psi^{\dag(0)}_{\alpha' k'\sigma}\equiv\sum_{l=0}^{\infty}\psi^{(l)}_{\alpha k\sigma}.
\end{align}
Using this expansion in Eq. \eqref{eq:eff_inter} we obtain
\begin{align}
{\cal H}&=\sum_{m=0}^{\infty}\sum_{\alpha=\pm 1}\beta_\alpha \left(\epsilon_k-\frac{\alpha\Phi}{2}\right)\sum_{m=0}^{l}\sum_{k\sigma}\psi^{\dag(k)}_{\alpha k\sigma}\psi^{(l-k)}_{\alpha k\sigma}\notag\\
&=\underbrace{\sum_{\alpha=\pm 1}\beta_\alpha \left(\epsilon_k-\frac{\alpha\Phi}{2}\right)\sum_{k\sigma}\psi^{\dag(0)}_{\alpha k\sigma}\psi^{(0)}_{\alpha k\sigma}}_{{\cal H}^{(0)}}+\sum_{m=1}^{\infty}\underbrace{\sum_{\alpha=\pm 1}\beta_\alpha \left(\epsilon_k-\frac{\alpha\Phi}{2}\right)\sum_{m=0}^{l}\sum_{k\sigma}\psi^{\dag(k)}_{\alpha k\sigma}\psi^{(l-k)}_{\alpha k\sigma}}_{{\cal H}^{(m)}}
\end{align}
Note : we are unable to use the traditional trick of time-slicing and introducing a functional integral since we have 2 different temperatures in the system. We recall the operator identity ($A$ and $B$ are general noncommuting operators)
\begin{align}
e^{-(A+B)}=e^{-A}\left[1+\sum_{m=1}^{\infty}(-1)^m\prod_{i=1}^{m}\left(\int_{0}^{x_{i -1}} dx_{i} e^{x_i A} B e^{-x_i A}\right)\right],
\end{align}
where we define the boundary condition $x_0=1$. The general perturbative expansion for an aribtrary product of operators, which we will denote by ${\cal O}$, is given by
\begin{align}
\langle{\cal O}\rangle&=\frac{\Tr\left[e^{-{\cal H}}{\cal O}\right]}{\Tr\left[e^{-{\cal H}}\right]}=\sum_{\nu=0}^{\infty}\frac{\Tr\left[e^{-{\cal H}^{(0)}}\left(1+\sum_{m=1}^{\infty}(-1)^m\prod_{i=1}^{m}\left(\int_{0}^{x_{i -1}} dx_{i} e^{x_i {\cal H}^{(0)}} {\cal H}_{\text{int}} e^{-x_i {\cal H}^{(0)}}\right)\right){\cal O}\right]}{\Tr\left[e^{-{\cal H}^{(0)}}\left(1+\sum_{m=1}^{\infty}(-1)^m\prod_{i=1}^{m}\left(\int_{0}^{x_{i -1}} dx_{i} e^{x_i {\cal H}^{(0)}} {\cal H}_{\text{int}} e^{-x_i {\cal H}^{(0)}}\right)\right)\right]},
\label{eq:expansion_2}
\end{align}
where we have defined
\begin{align}
{\cal H}_{\text{int}}=\sum_{m=1}^{\infty}{\cal H}^{(m)}.
\end{align}

From Eqs. \eqref{eq:expansion_1} and \eqref{eq:expansion_2} it is readily apparent that the powers of $H_{\text{int}}$ enter $G^{\text{ret}}_{d^{\dag}_{\sigma}d_{\sigma}}$ in 2 distinct ways : (1)\ the expansion of $D_{\sigma}$ and (2)\ the expansion of ${\cal H}$. For a second order computation
\begin{align}
G^{\text{ret}}_{d^{\dag}_{\sigma}d_{\sigma}}(\omega)&=\left\langle \left\{D_{\sigma},d^{\dag}_{\sigma}\right\}\right\rangle=\sum_{k=0}^{\infty}\left\langle \left\{D^{(k)}_{\sigma},d^{\dag}_{\sigma}\right\}\right\rangle=\underbrace{\sum_{k=0}^{2}\left\langle \left\{D^{(k)}_{\sigma},d^{\dag}_{\sigma}\right\}\right\rangle}_{\text{relevant}}+\underbrace{\sum_{k=3}^{\infty}\left\langle \left\{D^{(k)}_{\sigma},d^{\dag}_{\sigma}\right\}\right\rangle}_{\text{irrelevant}}.
\end{align}
We examine each of the relevant part above and isolate the contributions to ${\cal O} (H_{\text{int}}^2)$.

\subsubsection{$\left\langle \left\{D^{(0)}_{\sigma},d^{\dag}_{\sigma}\right\}\right\rangle$  contribution to ${\cal O} (H_{\text{int}}^2)$}
From Eq. \eqref{eq:expansion_1} we find
\begin{align}
\left\{D^{(0)}_{\sigma},d^{\dag}_{\sigma}\right\}&=\frac{t^{2}}{\Omega}\sum_{1}\frac{\left|g_{1}\right|^{2}}{\omega-\ep+i\eta}=g(\omega).
\end{align}
It immediately follows that
\begin{empheq}[box=\fbox]{align}
G^{\text{ret}(0)}_{d^{\dag}_{\sigma}d_{\sigma}}(\omega)=\left\langle \left\{D^{(0)}_{\sigma},d^{\dag}_{\sigma}\right\}\right\rangle=g(\omega)\left\langle \openone\right\rangle=g(\omega)=\frac{1}{\omega-\Delta+i\Gamma}.
\label{eq:D0}
\end{empheq}
Thus $\left\langle \left\{D^{(0)}_{\sigma},d^{\dag}_{\sigma}\right\}\right\rangle$ contributes to $G^{\text{ret}}_{d^{\dag}_{\sigma}d_{\sigma}}$ to 0th order in $H_{\text{int}}$ only.

\subsubsection{$\left\langle \left\{D^{(1)}_{\sigma},d^{\dag}_{\sigma}\right\}\right\rangle$  contribution to ${\cal O} (H_{\text{int}}^2)$}

First, we compute $D^{(1)}_{\sigma}(\omega)$ using Eq. \eqref{eq:recursion}
\begin{align}
D^{(1)}_{\sigma}(\omega)=-\frac{1}{\omega+\mathcal{ L}'+i\eta}\mathcal{ L_{\text I}}D^{(0)}_{\sigma}.
\label{eq:B2}
\end{align}
To compute this quantity we first evaluate the action of the superoperator $\mathcal{ L}_{\text I}$ on 
$\psi_{\bar{1}\sigma}$
\begin{align}
\mathcal{ L}_{\text I}\psi_{\bar{1}\sigma}
=-U&g^{\ast}_{\bar{1}}\left(\frac{t^{2}}{\Omega}\sum_{123}g^{\ast}_{1}g_{2}g_{3}\psi^{(0)\dag}_{1-\sigma}\psi^{(0)}_{2-\sigma}\psi^{(0)}_{3\sigma}-\frac{1}{2}\sum_{1}g_{1}\psi^{(0)}_{1\sigma}\right).
\end{align}
Using this expression in Eq. \eqref{eq:B2} we obtain
\begin{align}
D^{(1)}_{\sigma}(\omega)
&=U\frac{t}{\sqrt{\Omega}}g(\omega)\bigg(\frac{t^{2}}{\Omega}\sum_{123}\frac{g^{\ast}_{1}g_{2}g_{3}}{\omega+\ep_{1}-\ep_{2}-\ep_{3}+i\eta}\psi^{(0)\dag}_{1-\sigma}\psi^{(0)}_{2-\sigma}\psi^{(0)}_{3\sigma}-\frac{1}{2}\sum_{1}\frac{g_{1}}{\omega-\ep_{1}+i\eta}\psi^{(0)}_{1\sigma}\bigg).
\label{eq:d1}
\end{align}
The anticommutator of interest is then readily computed to give
\begin{align}
\left\{D^{(1)}_{\sigma}(\omega),d^{\dag}_{\sigma}\right\}=Ug(\omega)\left[-\frac{1}{2}g(\omega)+\frac{t^{2}}{\Omega}\sum_{12}g^{\ast}_{1}g_{2}g(\omega+\ep_{1}-\ep_{2})\psi^{(0)\dag}_{1-\sigma}\psi^{(0)}_{2-\sigma}\right].
\label{eq:A}
\end{align}
Now, this implies
\begin{align}
\left\langle \left\{D^{(1)}_{\sigma},d^{\dag}_{\sigma}\right\}\right\rangle=\sum_{\nu=0}^{\infty}\left\langle \left\{D^{(1)}_{\sigma},d^{\dag}_{\sigma}\right\}\right\rangle^{(\nu)}.
\end{align}
Note : We will use the superscript $\left\langle{\cal O}\right\rangle^{(n)}$ to denote the expectation value of the operator ${\cal O}$ evaluated with respect to the $n$'th order expansion of the exponent as defined above.
Thus,
\begin{empheq}[box=\fbox]{align}
G^{\text{ret}(1)}_{d^{\dag}_{\sigma}d_{\sigma}}(\omega)=\left\langle \left\{D^{(1)}_{\sigma},d^{\dag}_{\sigma}\right\}\right\rangle^{(0)}&=Ug(\omega)^{2}\bigg[\left(\frac{t^{2}}{\Omega}\right)\sum_{1}\left| g_{1}\right|^{2}f(\ep_{1})-\frac{1}{2}\bigg]=Ug(\omega)^{2}\bigg[\frac{\Gamma}{\pi}\int d\epsilon\left| g(\epsilon)\right|^{2}f^{\text{eff}}(\ep)-\frac{1}{2}\bigg].
\label{eq:1st_order_ret}
\end{empheq}
Here, we define the "effective" Fermi function
\begin{align}
f^{\text{eff}}(\epsilon)=\frac{1}{2}\left[\frac{1}{e^{\beta_1(\epsilon-\Phi/2)}+1}+\frac{1}{e^{\beta_{-1}(\epsilon+\Phi/2)}+1}\right].
\end{align}
The term $\left(\frac{\Gamma}{\pi}\int d\epsilon \left|g(\epsilon)\right|^2 f^{\text{eff}}(\epsilon)-\frac{1}{2}\right)$ vanishes when either $\Phi=0$ or $\beta_1=\beta_{-1}$ or both, but not if both $\Phi\neq 0$ and $\beta_1\neq\beta_{-1}$. For either purely thermal or bias-driven transport this contribution vanishes. Let us consider the first order correction to the Lippmann-Schwinger operator
\begin{align}
\psi^{\dag(1)}_{\alpha k\sigma}=U g(\epsilon_k)\left(\frac{t^2}{\Omega}\right)^2\sum_{\substack{\alpha_1 \alpha_2\alpha_3\\k_1 k_2 k_3}}\frac{g^{\ast}(\epsilon_1) g^{\ast}(\epsilon_2)g(\epsilon_3)}{\epsilon_k-\epsilon_1-\epsilon_2+\epsilon_3+i\eta}\psi^{\dag(0)}_{\alpha_1 k_1\sigma}\psi^{\dag(0)}_{\alpha_2 k_2 -\sigma}\psi^{(0)}_{\alpha_3 k_3 -\sigma}-\frac{U}{2}g(\epsilon_k)\frac{t^2}{\Omega}\sum_{\alpha_1 k_1}\frac{g^{\ast}(\epsilon_1)}{\epsilon_k-\epsilon_1+i\eta}\psi^{\dag(0)}_{\alpha_1 k_1\sigma}
\end{align}
 Thus,
\begin{align}
{\cal H}^{(1)}&=\sum_{\alpha k\sigma}\beta_{\alpha}\left(\epsilon_{k}-\alpha\frac{\Phi}{2}\right)\left(\psi^{\dag(1)}_{\alpha k\sigma}\psi^{(0)}_{\alpha k\sigma}+\psi^{\dag(0)}_{\alpha k\sigma}\psi^{(1)}_{\alpha k\sigma}\right)\notag\\
&=U\left(\frac{t^2}{\Omega}\right)^2\sum_{\substack{\alpha_1 \alpha_2\alpha_3\alpha_4\\k_1 k_2 k_3k_4\sigma}}\frac{g^{\ast}(\epsilon_1) g^{\ast}(\epsilon_2)g(\epsilon_3)g(\epsilon_4)}{\epsilon_3+\epsilon_4-\epsilon_1-\epsilon_2+i\eta}\left[\beta_{\alpha_4}\left(\epsilon_{4}-\alpha_4\frac{\Phi}{2}\right)-\beta_{\alpha_1}\left(\epsilon_{1}-\alpha_1\frac{\Phi}{2}\right)\right]\psi^{\dag(0)}_{\alpha_1 k_1\sigma}\psi^{\dag(0)}_{\alpha_2 k_2 -\sigma}\psi^{(0)}_{\alpha_3 k_3 -\sigma}\psi^{(0)}_{\alpha_4 k_4\sigma}\notag\\
&\qquad\qquad-\frac{U}{2}\frac{t^2}{\Omega}\sum_{\substack{\alpha_1 k_1\sigma\\\alpha_2 k_2}}\left[\beta_{\alpha_2}\left(\epsilon_{2}-\alpha_2\frac{\Phi}{2}\right)-\beta_{\alpha_1}\left(\epsilon_{1}-\alpha_1\frac{\Phi}{2}\right)\right] \frac{g^{\ast}(\epsilon_1)g(\epsilon_2)}{\epsilon_2-\epsilon_1+i\eta}\psi^{\dag(0)}_{\alpha_1 k_1\sigma}\psi^{(0)}_{\alpha_2 k_2 \sigma}.
\end{align}
It follows that
\begin{align}
\left\langle \underbrace{\left\{D^{(1)}_{\sigma},d^{\dag}_{\sigma}\right\}}_{{\cal O}}\right\rangle^{(1)}&=-\frac{\Tr\left[e^{-{\cal H}^{(0)}}\left(\int_{0}^{1} dx e^{x {\cal H}^{(0)}} {\cal H}^{(1)} e^{-x {\cal H}^{(0)}}\right){\cal O}\right]}{\Tr\left[e^{-{\cal H}^{(0)}}\right]}+\frac{\Tr\left[e^{-{\cal H}^{(0)}}\left(\int_{0}^{1} dx e^{x {\cal H}^{(0)}} {\cal H}^{(1)} e^{-x {\cal H}^{(0)}}\right)\right]}{\Tr\left[e^{-{\cal H}^{(0)}}\right]}\frac{\Tr\left[e^{-{\cal H}^{(0)}}{\cal O}\right]}{\Tr\left[e^{-{\cal H}^{(0)}}\right]},
\end{align}
Defining the interaction picture representation as ${\cal O}_I (x)=e^{x {\cal H}^{(0)}}{\cal O}e^{-x {\cal H}^{(0)}}$, and similarly for the other operators,  we can rewrite this as
\begin{align}
\left\langle {\cal O}\right\rangle^{(1)}&=-\frac{\Tr\left[e^{-{\cal H}^{(0)}}\int_{0}^{1} dx {\cal T}_{x}\left({\cal H}^{(1)}_{I}(x){\cal O}_I (0)\right)\right]}{\Tr\left[e^{-{\cal H}^{(0)}}\right]}+\frac{\Tr\left[e^{-{\cal H}^{(0)}}\int_{0}^{1} dx {\cal H}^{(1)}_{I}(x)\right]}{\Tr\left[e^{-{\cal H}^{(0)}}\right]}\frac{\Tr\left[e^{-{\cal H}^{(0)}}{\cal O}\right]}{\Tr\left[e^{-{\cal H}^{(0)}}\right]}.
\end{align}
We evaluate this expression by splitting up $\left\langle {\cal O}\right\rangle^{(1)}={\cal A}+{\cal B}$ as shown.
\begin{align}
{\cal A}&=\frac{U^2}{2}\left(\frac{t^2}{\Omega}\right)^2g(\omega)\sum_{\substack{\alpha_1 \alpha_2\alpha_3\alpha_4\\k_1 k_2 k_3 k_4\sigma'}}\left[\beta_{\alpha_2}\left(\epsilon_{2}-\alpha_2\frac{\Phi}{2}\right)-\beta_{\alpha_1}\left(\epsilon_{1}-\alpha_1\frac{\Phi}{2}\right)\right] \frac{g^{\ast}(\epsilon_1)g(\epsilon_2)}{\epsilon_2-\epsilon_1+i\eta}g^{\ast}(\epsilon_3)g(\epsilon_4)g(\omega+\epsilon_3-\epsilon_4)\notag\\
&\times\int_{0}^{1}dx\left\langle{\cal T}_x\left[
\contraction{}{\psi}{^{\dag(0)}_{\alpha_1 k_1\sigma'}(x)\psi^{(0)}_{\alpha_2 k_2 \sigma'}(x)\psi^{\dag(0)}_{\alpha_3 k_3-\sigma}(0)}{\psi}
\contraction[2ex]{\psi^{\dag(0)}_{\alpha_1 k_1\sigma'}(x)}{\psi}{^{(0)}_{\alpha_2 k_2 \sigma'}(x)}{\psi}
\psi^{\dag(0)}_{\alpha_1 k_1\sigma'}(x)\psi^{(0)}_{\alpha_2 k_2 \sigma'}(x)\psi^{\dag(0)}_{\alpha_3 k_3-\sigma}(0)\psi^{(0)}_{\alpha_4 k_4-\sigma}(0)
\right]\right\rangle.
\end{align}
Now, ${\cal H}^{(0)}$ is indeed quadratic and diagonal in the $\psi^{(0)}$ operators thus we can use Wick's theorem and obtain
\begin{align}
{\cal A}&=\frac{U^2}{2}\left(\frac{t^2}{\Omega}\right)^2g(\omega)\sum_{\substack{\alpha_1 \alpha_2\\k_1 k_2 }}\left[\beta_{\alpha_2}\left(\epsilon_{2}-\alpha_2\frac{\Phi}{2}\right)-\beta_{\alpha_1}\left(\epsilon_{1}-\alpha_1\frac{\Phi}{2}\right)\right]\frac{\left| g(\epsilon_1)\right|^2\left|g(\epsilon_2)\right|^2}{\epsilon_2-\epsilon_1+i\eta} g(\omega+\epsilon_2-\epsilon_1)\notag\\
&\qquad\qquad\qquad\qquad\times\int_{0}^{1}dx
\left\langle{\cal T}_x\left[\psi^{\dag(0)}_{\alpha_1 k_1-\sigma}(x)\psi^{(0)}_{\alpha_1 k_1-\sigma}(0)\right]\right\rangle 
\left\langle{\cal T}_x\left[\psi^{(0)}_{\alpha_2 k_2-\sigma}(x)\psi^{\dag(0)}_{\alpha_2 k_2-\sigma}(0)\right]\right\rangle\notag\\
&=\frac{U^2}{2}\left(\frac{t^2}{\Omega}\right)^2g(\omega)\sum_{\substack{\alpha_1 \alpha_2\\k_1 k_2 }}\frac{\left| g(\epsilon_1)\right|^2\left|g(\epsilon_2)\right|^2}{\epsilon_2-\epsilon_1+i\eta} g(\omega+\epsilon_2-\epsilon_1)\left[f_{\alpha_1}(\epsilon_1)-f_{\alpha_2}(\epsilon_2)\right]
\label{eq:1st_partA}
\end{align}
%

Similarly for the other part
\begin{align}
{\cal B}&=U^2\left(\frac{t^2}{\Omega}\right)^3g(\omega)\sum_{\substack{\alpha_1 \alpha_2\alpha_3\alpha_4\\\alpha_5\alpha_6 k_1 k_2 k_3\\k_4 k_5 k_6\sigma'}}\left[\tilde{\epsilon}_{\alpha_1 k_1}-\tilde{\epsilon}_{\alpha_4 k_4}\right] \frac{g^{\ast}(\epsilon_1)g^{\ast}(\epsilon_2)g(\epsilon_3)g(\epsilon_4)}{\epsilon_3+\epsilon_4-\epsilon_1-\epsilon_2+i\eta}g^{\ast}(\epsilon_5)g(\epsilon_6)g(\omega+\epsilon_5-\epsilon_6)\notag\\
&\times\int_{0}^{1}dx\left\langle{\cal T}_x\left[
\contraction{}{\psi}{^{\dag(0)}_{\alpha_1 k_1\sigma'}(x)\psi^{\dag(0)}_{\alpha_2 k_2-\sigma'}(x)\psi^{(0)}_{\alpha_3 k_3 -\sigma'}(x)\psi^{(0)}_{\alpha_4 k_4 \sigma'}(x)\psi^{\dag(0)}_{\alpha_5 k_5-\sigma}(0)}{\psi}
\contraction[2ex]{\psi^{\dag(0)}_{\alpha_1 k_1\sigma'}(x)}{\psi}{^{\dag(0)}_{\alpha_2 k_2-\sigma'}(x)}{\psi}
\contraction[2ex]{\psi^{\dag(0)}_{\alpha_1 k_1\sigma'}(x)\psi^{\dag(0)}_{\alpha_2 k_2\sigma'}(x)\psi^{(0)}_{\alpha_3 k_3 -\sigma'}(x)}{\psi}{^{(0)}_{\alpha_4 k_4 \sigma'}(x)}{\psi}
\bcontraction{}{\psi}{^{\dag(0)}_{\alpha_1 k_1\sigma'}(x)\psi^{\dag(0)}_{\alpha_2 k_2-\sigma'}(x)\psi^{(0)}_{\alpha_3 k_3 -\sigma'}(x)}{\psi}
\bcontraction[2ex]{\psi^{\dag(0)}_{\alpha_1 k_1\sigma'}(x)}{\psi}{^{\dag(0)}_{\alpha_2 k_2-\sigma'}(x)\psi^{(0)}_{\alpha_3 k_3 -\sigma'}(x)\psi^{(0)}_{\alpha_4 k_4 \sigma'}(x)\psi^{\dag(0)}_{\alpha_5 k_5-\sigma}(0)}{\psi}
\bcontraction[3ex]{\psi^{\dag(0)}_{\alpha_1 k_1\sigma'}(x)\psi^{\dag(0)}_{\alpha_2 k_2-\sigma'}(x)}{\psi}{^{(0)}_{\alpha_3 k_3 -\sigma'}(x)\psi^{(0)}_{\alpha_4 k_4 \sigma'}(x)}{\psi}
\psi^{\dag(0)}_{\alpha_1 k_1\sigma'}(x)\psi^{\dag(0)}_{\alpha_2 k_2-\sigma'}(x)\psi^{(0)}_{\alpha_3 k_3 -\sigma'}(x)\psi^{(0)}_{\alpha_4 k_4 \sigma'}(x)\psi^{\dag(0)}_{\alpha_5 k_5-\sigma}(0)\psi^{(0)}_{\alpha_6 k_6-\sigma}(0)
\right]\right\rangle.
\end{align}
The lower contraction vanishes when we contract  $\psi^{\dag(0)}_{\alpha_1 k_1\sigma'}$ and $\psi^{\dag(0)}_{\alpha_4 k_4\sigma'}$ and we obtain
\begin{align}
{\cal B}&=U^2\left(\frac{t^2}{\Omega}\right)^3g(\omega)\sum_{\substack{\alpha_1 \alpha_2 \alpha_3\\ k_1 k_2 k_3}}\left[\tilde{\epsilon}_{\alpha_3 k_3}-\tilde{\epsilon}_{\alpha_1 k_1}\right]\frac{\left| g(\epsilon_1)\right|^2\left| g(\epsilon_2)\right|^2\left| g(\epsilon_3)\right|^2}{\epsilon_3-\epsilon_1+i\eta}g(\omega+\epsilon_3-\epsilon_1)\notag\\
&\times\int_{0}^{1}dx\left\langle{\cal T}_x\left[\psi^{\dag(0)}_{\alpha_1 k_1-\sigma}(x)\psi^{(0)}_{\alpha_1 k_1-\sigma}(0)\right]\right\rangle \left\langle{\cal T}_x\left[\psi^{\dag(0)}_{\alpha_2 k_2\sigma}(x)\psi^{(0)}_{\alpha_2 k_2\sigma}(x)\right]\right\rangle\left\langle{\cal T}_x\left[\psi^{\dag(0)}_{\alpha_3 k_3-\sigma}(0)\psi^{(0)}_{\alpha_3 k_3-\sigma}(x)\right]\right\rangle\notag\\
&=U^2 g(\omega)\left[\frac{t^2}{\Omega}\sum_{\alpha k} \left|g(\epsilon)\right|^2 f^{\text{eff}}(\epsilon)\right]\left[\left(\frac{t^2}{\Omega}\right)^2\sum_{\substack{\alpha_1 \alpha_2\\k_1 k_2}}\frac{\left| g(\epsilon_1)\right|^2\left| g(\epsilon_2)\right|^2}{\epsilon_2-\epsilon_1+i\eta}g(\omega+\epsilon_2-\epsilon_1)\left[f_{\alpha_2}(\epsilon_2)-f_{\alpha_1}(\epsilon_1)\right]\right].
\label{eq:1st_partB_2}
\end{align}
Collecting both parts in Eqs. \eqref{eq:1st_partA} and \eqref{eq:1st_partB_2} we have
\begin{empheq}[box=\fbox]{align}
\left\langle \left\{D^{(1)}_{\sigma},d^{\dag}_{\sigma}\right\}\right\rangle^{(1)}&=U^2 g(\omega)\left[\frac{\Gamma}{\pi}\int_{-\infty}^{\infty} d\epsilon \left|g(\epsilon)\right|^2 f^{\text{eff}}(\epsilon)-\frac{1}{2}\right]\notag\\
&\times\bigg[\frac{\Gamma}{\pi}\int d\epsilon'\left| g(\epsilon')\right|^2 f^{\text{eff}}(\epsilon')\left[\frac{1}{(\epsilon'-\Delta+i\Gamma)(\epsilon'+\omega-2\Delta+i2\Gamma)}-\frac{1}{(\epsilon'-\Delta-i\Gamma)(\epsilon'-\omega-i2\Gamma)}\right].
\label{eq:1st_partB}
\end{empheq}
We observe when $\Delta=0$ setting either $\Phi=0$ or $\beta_1=\beta_{-1}$, {\it i.e.} for purely bias-driven or thermal driven transport (or the trivial case of both), $\frac{\Gamma}{\pi}\int_{-\infty}^{\infty} d\epsilon \left|g(\epsilon)\right|^2 f^{\text{eff}}(\epsilon)=1/2$ and this contribution vanishes.

\subsubsection{$\left\langle \left\{D^{(2)}_{\sigma},d^{\dag}_{\sigma}\right\}\right\rangle$  contribution to ${\cal O} (H_{\text{int}}^2)$}

Similarly $D^{(2)}_{\sigma}(\omega)$ follows from the use of Eq.\ \eqref{eq:recursion} in Eq.\ \eqref{eq:d1} above
\begin{align}
D^{(2)}_{\sigma}(\omega)=-\frac{1}{\omega+\mathcal{ L}'+i\eta}\mathcal{ L_{\text I}}D^{(1)}_{\sigma}.
\end{align}

\noindent We begin by computing 
\begin{align}
\mathcal{ L_{\text I}}D^{(1)}_{\sigma}&=\frac{U}{2}\bigg[-\frac{t^{2}}{\Omega}\sum_{12\bar{\sigma}}g^{\ast}_{1}g_{2}\psi^{(0)\dag}_{1\bar{\sigma}}\psi^{(0)}_{2\bar{\sigma}}+\left(\frac{t^{2}}{\Omega}\right)^{2}\sum_{1234\bar{\sigma}}g^{\ast}_{1}g^{\ast}_{2}g_{3}g_{4}\psi^{(0)\dag}_{1\bar{\sigma}}\psi^{(0)\dag}_{2-\bar{\sigma}}\psi^{(0)}_{3-\bar{\sigma}}\psi^{(0)}_{4\bar{\sigma}},D^{(1)}_{\sigma}\bigg]\notag\\
&=-\frac{U}{2}\bigg[\frac{t^{2}}{\Omega}\sum_{12\bar{\sigma}}g^{\ast}_{1}g_{2}\psi^{(0)\dag}_{1\bar{\sigma}}\psi^{(0)}_{2\bar{\sigma}},D^{(1)}_{\sigma}\bigg]+\frac{U}{2}\bigg[\left(\frac{t^{2}}{\Omega}\right)^{2}\sum_{1234\bar{\sigma}}g^{\ast}_{1}g^{\ast}_{2}g_{3}g_{4}\psi^{(0)\dag}_{1\bar{\sigma}}\psi^{(0)\dag}_{2-\bar{\sigma}}\psi^{(0)}_{3-\bar{\sigma}}\psi^{(0)}_{4\bar{\sigma}},D^{(1)}_{\sigma}\bigg].
\label{eq:temp}
\end{align}

It is convenient to divide the action of the Liouvillian above into two parts by defining 
\begin{align}
&\bar{\mathcal{A}}=-\frac{U}{2}\bigg[\frac{t^{2}}{\Omega}\sum_{12\bar{\sigma}}g^{\ast}_{1}g_{2}\psi^{(0)\dag}_{1\bar{\sigma}}\psi^{(0)}_{2\bar{\sigma}},D^{(1)}_{\sigma}\bigg]\notag\\
&\bar{\mathcal{B}}=\frac{U}{2}\bigg[\left(\frac{t^{2}}{\Omega}\right)^{2}\sum_{1234\bar{\sigma}}g^{\ast}_{1}g^{\ast}_{2}g_{3}g_{4}\psi^{(0)\dag}_{1\bar{\sigma}}\psi^{(0)\dag}_{2-\bar{\sigma}}\psi^{(0)}_{3-\bar{\sigma}}\psi^{(0)}_{4\bar{\sigma}},D^{(1)}_{\sigma}\bigg].
\end{align}

This can be simplified to give
\begin{align}
\bar{\mathcal{A}}=&\frac{U^2}{2}
\left(\frac{t}{\sqrt{\Omega}}\right)^{3/2}g(\omega)\sum_{123}g^{\ast}g_2 g_3\left[g(\omega+\epsilon_1-\epsilon_2)+g(\omega+\epsilon_1-\epsilon_3)-g(\omega-\epsilon_2-\epsilon_3)\right]\psi^{\dag}_{1-\sigma}\psi_{2-\sigma}\psi_{3\sigma}\notag\\
&-\frac{U^2}{4}\frac{t}{\sqrt{\Omega}}\left[g(\omega)\right]^2\sum_{1}g_1\psi_{1\sigma}
\end{align}
\noindent We introduce the notation
\begin{align}
&D^{(2)}_{\sigma\bar{\mathcal{ (A)}}}(\omega)=-\frac{1}{\omega+\mathcal{ L}'+i\eta}\mathcal{ L_{\text I}}\bar{\mathcal{A}} \notag\\
&D^{(2)}_{\sigma\bar{\mathcal{ (B)}}}(\omega)=-\frac{1}{\omega+\mathcal{ L}'+i\eta}\mathcal{ L_{\text I}}\bar{\mathcal{B}},
\end{align}
to separate the contributions to $D^{(2)}_{\sigma}(\omega)$ which follow from the parts $\bar{\mathcal{A}}$ and $\bar{\mathcal{B}}$.
Simplifying part $\bar{{\cal A}}$ we get
\begin{align}
D^{(2)}_{\sigma(\bar{\mathcal{A})}}(\omega)&=\frac{U^{2}}{2}g(\omega)\frac{t}{\sqrt{\Omega}}\bigg[\frac{t^{2}}{\Omega}\sum_{123}\frac{g^{\ast}_{1}g_{2}g_{3}}{\omega+\ep_{1}-\ep_{2}-\ep_{3}+i\eta}\left(g(\omega-\ep_{2}-\ep_{3})-g(\omega+\ep_{1}-\ep_{2})-g(\omega+\ep_{1}-\ep_{3})\right)\psi^{\dag}_{1-\sigma}\psi_{2-\sigma}\psi_{3\sigma}\notag\\
&\qquad\qquad\qquad\qquad\qquad\qquad+\frac{g(\omega)}{2}\sum_{1}\frac{g_{1}}{\omega-\ep_{1}+i\eta}\psi_{1\sigma}\bigg].
\end{align}
Evaluating the anticommutator we obtain
\begin{align}
\left\{D^{(2)}_{\sigma(\bar{\mathcal{A}})}(\omega),d^{\dag}_{\sigma}\right\}&=\frac{t}{\sqrt{\Omega}}\sum_{\bar{1}}g_{\bar{1}}\left\{D^{(2)}_{\sigma(\bar{\mathcal{A}})},\psi^{\dag}_{\bar{1}\sigma}\right\}\notag\\
&=\frac{U^{2}}{2}g(\omega)\frac{t^{2}}{\Omega}\bigg[\frac{t^{2}}{\Omega}\sum_{123}\frac{g^{\ast}_{1}g_{2}\left|g_{3}\right|^{2}}{\omega+\ep_{1}-\ep_{2}-\ep_{3}+i\eta}\left(g(\omega-\ep_{2}-\ep_{3})-g(\omega+\ep_{1}-\ep_{2})-g(\omega+\ep_{1}-\ep_{3})\right)\psi^{\dag}_{1-\sigma}\psi_{2-\sigma}\notag\\
&\qquad\qquad\qquad\qquad\qquad\qquad+\frac{g(\omega)}{2}\sum_{1}\frac{\left|g_{1}\right|^{2}}{\omega-\ep_{1}+i\eta}\bigg].
\end{align}

\noindent The evaluation of the Green's function is likewise separated into two parts. First, we compute the contribution of the part $\bar{\mathcal{A}}$  
\begin{empheq}[box=\fbox]{align}
&G^{\text{ret}(2)}_{d_{\sigma}d^{\dag}_{\sigma}(\bar{\mathcal{A}})}(\omega)=\left\langle\left\{D^{(2)}_{\sigma(\bar{\mathcal{A}})}(\omega),d^{\dag}_{\sigma}\right\}\right\rangle^{(0)}\notag\\
&=\frac{U^{2}}{2}\left[g(\omega)\right]^{2}\bigg[\frac{\Gamma}{\pi}\int{d\ep_{1}}\left|g_{1}\right|^{2}\left(\frac{1}{\omega-\ep_{1}-2\Delta+2i\Gamma}-\frac{1}{\omega+\ep_{1}-2\Delta+2i\Gamma}\right)f^{\text{eff}}_{1}-g(\omega)\left(\frac{\Gamma}{\pi}\int d\epsilon \left|g(\epsilon)\right|^2 f^{\text{eff}}(\epsilon)-\frac{1}{2}\right)\bigg].
\label{eq:bitA}
\end{empheq}
Once again, at the particle-hole symmetric point the term $\left(\frac{\Gamma}{\pi}\int d\epsilon g(\epsilon) f^{\text{eff}}(\epsilon)-\frac{1}{2}\right)$ vanishes when either $\Phi=0$ or $\beta_1=\beta_{-1}$ or both, but not if both $\Phi\neq 0$ and $\beta_1\neq\beta_{-1}$. For either purely thermal or bias-driven transport at the point $\epsilon_d=-U/2$ we get
\begin{align}
&G^{\text{ret}(2)}_{d_{\sigma}d^{\dag}_{\sigma}(\bar{\mathcal{A}})}(\omega)=\left\langle\left\{D^{(2)}_{\sigma(\bar{\mathcal{A}})}(\omega),d^{\dag}_{\sigma}\right\}\right\rangle^{(0)}\notag\\
&\quad=\frac{U^{2}}{2}\left[g(\omega)\right]^{2}\bigg[\frac{\Gamma}{\pi}\int{d\ep_{1}}\left|g_{1}\right|^{2}\left(\frac{1}{\omega-\ep_{1}+2i\Gamma}-\frac{1}{\omega+\ep_{1}+2i\Gamma}\right)f^{\text{eff}}_{1}\bigg].
\label{eq:bitA_simplified}
\end{align}
\noindent Similarly, we compute the the contribution of part $\bar{\mathcal{B}}$
\begin{align}
D^{(2)}_{\sigma(\bar{\mathcal{B}})}(\omega)=&-U^{2}g(\omega)\frac{t}{\sqrt{\Omega}}\frac{t^{2}}{\Omega}\bigg[\frac{t^{2}}{\Omega}\sum_{12345}\frac{g^{\ast}_{1}g^{\ast}_{2}g_{3}g_{4}g_{5}}{\omega+\ep_{1}+\ep_{2}-\ep_{3}-\ep_{4}-\ep_{5}+i\eta}\left(g(\omega-\ep_{3}-\ep_{5})+g(\omega+\ep_{2}-\ep_{4})\right)\psi^{\dag}_{1\sigma}\psi^{\dag}_{2-\sigma}\psi_{3-\sigma}\psi_{4\sigma}\psi_{5\sigma}\notag\\
&\qquad\qquad\qquad\qquad+\frac{t^{2}}{\Omega}\sum_{12345}\frac{g^{\ast}_{1}g^{\ast}_{2}g_{3}g_{4}g_{5}}{\omega+\ep_{1}+\ep_{2}-\ep_{3}-\ep_{4}-\ep_{5}+i\eta}g(\omega+\ep_{1}-\ep_{3})\psi^{\dag}_{1-\sigma}\psi^{\dag}_{2-\sigma}\psi_{3-\sigma}\psi_{4-\sigma}\psi_{5\sigma}\notag\\
&\qquad\qquad+\sum_{123}\frac{g^{\ast}_{1}g_{2}g_{3}}{\omega+\ep_{1}-\ep_{2}-\ep_{3}+i\eta}\left(\frac{g(\omega)}{2}-\frac{t^2}{\Omega}\underbrace{\sum_{4}\left|g(\epsilon_4)\right|^2 g(\omega+\epsilon_1-\epsilon_4)}_{\omega+\epsilon_1-2\Delta+2i\Gamma}\right)\psi^{\dag}_{1-\sigma}\psi^{\dag}_{2-\sigma}\psi_{3\sigma}\bigg].
\end{align}
Evaluating the anticommutator we get
\begin{align}
\left\{D^{(2)}_{\sigma(\bar{\mathcal{B}})}(\omega),d^{\dag}_{\sigma}\right\}&=\frac{t}{\sqrt{\Omega}}\sum_{\bar{1}}g_{\bar{1}}\left\{D^{2}_{\sigma(\text{I})}(\omega),\psi^{\dag}_{\bar{1}\sigma}\right\}\notag\\
=-U^{2}&g(\omega)\left(\frac{t^{2}}{\Omega}\right)^{2}\bigg[\frac{t^{2}}{\Omega}\sum_{12345}\frac{g^{\ast}_{1}g^{\ast}_{2}g_{3}g_{4}\left|g_{5}\right|^{2}}{\omega+\ep_{1}+\ep_{2}-\ep_{3}-\ep_{4}-\ep_{5}+i\eta}\left(g(\omega-\ep_{3}-\ep_{5})+g(\omega+\ep_{2}-\ep_{4})\right)\psi^{\dag}_{1\sigma}\psi^{\dag}_{2-\sigma}\psi_{3-\sigma}\psi_{4\sigma}\notag\\
&\qquad\qquad-\frac{t^{2}}{\Omega}\sum_{12345}\frac{g^{\ast}_{1}g^{\ast}_{2}g_{3}\left|g_{4}\right|^{2}g_{5}}{\omega+\ep_{1}+\ep_{2}-\ep_{3}-\ep_{4}-\ep_{5}+i\eta}\left(g(\omega-\ep_{3}-\ep_{5})+g(\omega+\ep_{2}-\ep_{4})\right)\psi^{\dag}_{1\sigma}\psi^{\dag}_{2-\sigma}\psi_{3-\sigma}\psi_{5\sigma}\notag\\
&\qquad\qquad\qquad\qquad+\frac{t^{2}}{\Omega}\sum_{12345}\frac{g^{\ast}_{1}g^{\ast}_{2}g_{3}g_{4}\left|g_{5}\right|^{2}}{\omega+\ep_{1}+\ep_{2}-\ep_{3}-\ep_{4}-\ep_{5}+i\eta}g(\omega+\ep_{1}-\ep_{3})\psi^{\dag}_{1-\sigma}\psi^{\dag}_{2-\sigma}\psi_{3-\sigma}\psi_{4-\sigma}\notag\\
&\qquad\qquad\qquad\qquad\qquad\qquad+\sum_{123}\frac{g^{\ast}_{1}g_{2}\left|g_{3}\right|^{2}}{\omega+\ep_{1}-\ep_{2}-\ep_{3}+i\eta}\left(\frac{g(\omega)}{2}-\frac{1}{\omega+\ep_{1}-2\Delta+2i\Gamma}\right)\psi^{\dag}_{1-\sigma}\psi_{2-\sigma}\bigg].
\end{align}
Its contribution to the retarded electron Green's function of the dot $G^{\text{ret}(2)}_{d_{\sigma}d^{\dag}_{\sigma}}(\omega)$
\begin{align}
G^{\text{ret}(2)}_{d_{\sigma}d^{\dag}_{\sigma}((\bar{\mathcal{B}})}(\omega)=&\left\langle\left\{D^{(2)}_{\sigma(\bar{\mathcal{B}})}(\omega),d^{\dag}_{\sigma}\right\}\right\rangle^{(0)}\notag\\
=-U^{2}g(\omega)\left(\frac{t^{2}}{\Omega}\right)^{2}&\bigg[\frac{t^{2}}{\Omega}\sum_{12345}\frac{g^{\ast}_{1}g^{\ast}_{2}g_{3}g_{4}\left|g_{5}\right|^{2}}{\omega+\ep_{1}+\ep_{2}-\ep_{3}-\ep_{4}-\ep_{5}+i\eta}\left(g(\omega-\ep_{3}-\ep_{5})+g(\omega+\ep_{2}-\ep_{4})\right)\left\langle \contraction{}{\psi}{^{\dag}_{1\sigma}\psi^{\dag}_{2-\sigma}\psi_{3-\sigma}}{\psi}
\contraction[2ex]{\psi^{\dag}_{1\sigma}}{\psi}{^{\dag}_{2-\sigma}}{\psi}
\psi^{\dag}_{1\sigma}\psi^{\dag}_{2-\sigma}\psi_{3-\sigma}\psi_{4\sigma}
\right\rangle^{(0)}\notag\\
&-\frac{t^{2}}{\Omega}\sum_{12345}\frac{g^{\ast}_{1}g^{\ast}_{2}g_{3}\left|g_{4}\right|^{2}g_{5}}{\omega+\ep_{1}+\ep_{2}-\ep_{3}-\ep_{4}-\ep_{5}+i\eta}\left(g(\omega-\ep_{3}-\ep_{5})+g(\omega+\ep_{2}-\ep_{4})\right)\left\langle \contraction{}{\psi}{^{\dag}_{1\sigma}\psi^{\dag}_{2-\sigma}\psi_{3-\sigma}}{\psi}
\contraction[2ex]{\psi^{\dag}_{1\sigma}}{\psi}{^{\dag}_{2-\sigma}}{\psi}
\psi^{\dag}_{1\sigma}\psi^{\dag}_{2-\sigma}\psi_{3-\sigma}\psi_{5\sigma}
\right\rangle^{(0)}\notag\\
&\qquad\qquad\qquad+\frac{t^{2}
}{\Omega}\sum_{12345}\frac{g^{\ast}_{1}g^{\ast}_{2}g_{3}g_{4}\left|g_{5}\right|^{2}}{\omega+\ep_{1}+\ep_{2}-\ep_{3}-\ep_{4}-\ep_{5}+i\eta}g(\omega+\ep_{1}-\ep_{3})\left\langle 
\contraction{}{\psi}{^{\dag}_{1-\sigma}\psi^{\dag}_{2-\sigma}}{\psi}
\contraction[2ex]{\psi^{\dag}_{1-\sigma}}{\psi}{^{\dag}_{2-\sigma}\psi_{3-\sigma}}{\psi}
\bcontraction{}{\psi}{^{\dag}_{1-\sigma}\psi^{\dag}_{2-\sigma}\psi_{3-\sigma}}{\psi}
\bcontraction[2ex]{\psi^{\dag}_{1-\sigma}}{\psi}{^{\dag}_{2-\sigma}}{\psi}
\psi^{\dag}_{1-\sigma}\psi^{\dag}_{2-\sigma}\psi_{3-\sigma}\psi_{4-\sigma}\right\rangle^{(0)}\notag\\
&\qquad\qquad\qquad\qquad\qquad+\sum_{123}\frac{g^{\ast}_{1}g_{2}\left|g_{3}\right|^{2}}{\omega+\ep_{1}-\ep_{2}-\ep_{3}+i\eta}\left(\frac{g(\omega)}{2}-\frac{1}{\omega+\ep_{1}-2\Delta+2i\Gamma}\right)\left\langle 
\contraction{}{\psi}{^{\dag}_{1-\sigma}}{\psi}
\psi^{\dag}_{1-\sigma}\psi_{2-\sigma}\right\rangle^{(0)}\bigg].
\end{align}
We rewrite this in a simplified form without making general assumptions
\begin{empheq}[box=\fbox]{align}
G^{\text{ret}(2)}_{d_{\sigma}d^{\dag}_{\sigma}(\mathcal{B})}(\omega)&=-U^2[g(\omega)]^2\bigg[\left(\frac{\Gamma}{\pi}\right)^2\int d\epsilon_1 d\epsilon_2 \left|g_1\right|^2\left|g_2\right|^2 f_{1}^{\text{eff}}f_{2}^{\text{eff}}\bigg(\frac{1}{\omega-\epsilon_2-2\Delta+2i\Gamma}+2g(\omega+\epsilon_1-\epsilon_2)\notag\\
&-g(\omega-\epsilon_1-\epsilon_2)-g(\omega)-\frac{1}{\omega+\epsilon_2-2\Delta+2i\Gamma}\bigg)+\frac{\Gamma}{\pi}\int d\epsilon\left|g(\epsilon)\right|^2 f^{\text{eff}}(\epsilon)\left(\frac{g(\omega)}{2}-\frac{1}{\omega+\epsilon-2\Delta+2 i\Gamma}\right) \bigg].
\label{eq:bitB}
\end{empheq}
Let us now assume that either $\Phi=0$ or $T_1=T_{-1}$ (but not necessarily both), and that $\Delta=0$. In this situation we have the identity $f(-\epsilon)^{\text{eff}}=1-f(\epsilon)^{\text{eff}}$ which leads to the fact that $\frac{\Gamma}{\pi}\int d\epsilon \left|g(\epsilon)\right|^2 f(\epsilon)^{\text{eff}}=\frac{1}{2}$. This allows us to simplify this term which gives
\begin{align}
&G^{\text{ret}(2)}_{d_{\sigma}d^{\dag}_{\sigma}(\bar{\mathcal{B}})}(\omega)=-U^{2}\left[g(\omega)\right]^{2}\bigg[3 \left(\frac{\Gamma}{\pi}\right)^{2}\int{d\ep_{1}}\int{d\ep_{2}}\left|g_{1}\right|^{2}\left|g_{2}\right|^{2}g(\omega+\ep_{1}-\ep_{2})f_{1}^{\text{eff}}f_{2}^{\text{eff}}\notag\\
&\qquad\qquad\qquad\qquad-\frac{1}{2}\frac{1}{\omega+3i\Gamma}-\frac{\Gamma}{\pi}\int{d\ep_{1}}\left|g_{1}\right|^{2}\left(\frac{1}{\omega+\ep_{1}+2i\Gamma}\right)f_{1}^{\text{eff}}\bigg].
\label{eq:bitB_simplified}
\end{align}

\subsection{Evaluation of the retarded self-energy}\label{Appendix_final}
The Dyson equation 
\begin{align}
G^{\text{ret}}_{d_{\sigma}d^{\dag}_{\sigma}}(\omega)=G^{\text{ret}(0)}_{d_{\sigma}d^{\dag}_{\sigma}}(\omega)+G^{\text{ret}(0)}_{d_{\sigma}d^{\dag}_{\sigma}}(\omega)\Sigma(\omega)G^{\text{ret}(0)}_{d_{\sigma}d^{\dag}_{\sigma}}(\omega),
\label{eq:Dyson}
\end{align}
allows us to compute the $n$-th order ($n\geq 1$) contribution in U to the self energy of the system via the relation
\begin{align}
\Sigma^{(n)}=\frac{1}{g(\omega)^2}G^{\text{ret}(n)}_{d_{\sigma}d^{\dag}_{\sigma}}(\omega)
\label{eq:self_energy_pert}
\end{align}
Here $\Sigma(\omega)$ denotes the {\bf complete} self energy of the system. However for the Born Approximation we need to evaluate the {\bf proper} self energy of the system $\Sigma^{\star}$ which satisfies
\begin{align}
G^{\text{ret}}_{d_{\sigma}d^{\dag}_{\sigma}}(\omega)=G^{\text{ret}(0)}_{d_{\sigma}d^{\dag}_{\sigma}}(\omega)+G^{\text{ret}(0)}_{d_{\sigma}d^{\dag}_{\sigma}}(\omega)\Sigma^{\star}(\omega)G^{\text{ret}}_{d_{\sigma}d^{\dag}_{\sigma}}(\omega).
\label{eq:Dyson_proper}
\end{align}

Using Eq.\ \eqref{eq:self_energy_pert} in Eq.\ \eqref{eq:1st_order_ret} we get
\begin{align}
\Sigma^{(1)}=U\bigg[\frac{\Gamma}{\pi}\int d\epsilon\left| g(\epsilon)\right|^{2}f^{\text{eff}}(\ep)-\frac{1}{2}\bigg].
\label{eq:self_1st}
\end{align}
We collect the contributions to $\Sigma^{(2)}$ from Eqs.\ \eqref{eq:1st_partB},\eqref{eq:bitA} and \eqref{eq:bitB} which we call $\Sigma^{(2)}_{I}$,$\Sigma^{(2)}_{II}$ and $\Sigma^{(2)}_{III}$ respectively.
\begin{align}
\Sigma^{(2)}_{I}&=U^2\left[\frac{\Gamma}{\pi}\int_{-\infty}^{\infty} d\epsilon \left|g(\epsilon)\right|^2 f^{\text{eff}}(\epsilon)-\frac{1}{2}\right]\notag\\
&\times\bigg[\frac{\Gamma}{\pi}\int d\epsilon'\left| g(\epsilon')\right|^2 f^{\text{eff}}(\epsilon')\left(\frac{1}{(\epsilon'-\Delta+i\Gamma)}+\frac{1}{(\epsilon'-\Delta-i\Gamma)}-\frac{1}{(\epsilon'+\omega-2\Delta+i2\Gamma)}-\frac{1}{(\epsilon'-\omega-i2\Gamma)}\right)\bigg]
\end{align}
\begin{align}
\Sigma^{(2)}_{II}&=\frac{U^{2}}{2}\bigg[\frac{\Gamma}{\pi}\int{d\ep}\left|g(\ep)\right|^{2}\left(\frac{1}{\omega-\ep-2\Delta+2i\Gamma}-\frac{1}{\omega+\ep-2\Delta+2i\Gamma}\right)f^{\text{eff}}(\ep)-g(\omega)\left(\frac{\Gamma}{\pi}\int d\epsilon g(\epsilon) f^{\text{eff}}(\epsilon)-\frac{1}{2}\right)\bigg].
\label{eq:not_enough_1}
\end{align}
\begin{align}
\Sigma^{(2)}_{III}&=-U^2\bigg[\left(\frac{\Gamma}{\pi}\right)^2\int d\epsilon_1 d\epsilon_2 \left|g_1\right|^2\left|g_2\right|^2 f_{1}^{\text{eff}}f_{2}^{\text{eff}}\bigg(\frac{1}{\omega-\epsilon_2-2\Delta+2i\Gamma}+2g(\omega+\epsilon_1-\epsilon_2)-g(\omega-\epsilon_1-\epsilon_2)\notag\\
&-g(\omega)-\frac{1}{\omega+\epsilon_2-2\Delta+2i\Gamma}\bigg)+\frac{\Gamma}{\pi}\int d\epsilon\left|g(\epsilon)\right|^2 f^{\text{eff}}(\epsilon)\left(\frac{g(\omega)}{2}-\frac{1}{\omega+\epsilon-2\Delta+2 i\Gamma}\right) \bigg].
\label{eq:not_enough_2}
\end{align}
Collecting the contributions from Eq.\ \eqref{eq:not_enough_1} and \eqref{eq:not_enough_2} we can simplify it to obtain
\begin{align}
&\Sigma^{(2)}_{II+III}=U^2\bigg[\frac{\Gamma}{\pi}\int d\epsilon \left|g(\epsilon)\right|^2 f^{\text{eff}}(\epsilon)\bigg(\frac{1}{2}\left(\frac{1}{\omega-\epsilon-2\Delta+2i\Gamma}+\frac{1}{\omega+\epsilon-2\Delta+2i\Gamma}\right)-g(\omega)\bigg)+\frac{g(\omega)}{4}\notag\\
&+\left(\frac{\Gamma}{\pi}\right)^2\int d\epsilon_1 d\epsilon_2 \left|g_1\right|^2\left|g_2\right|^2 f_{1}^{\text{eff}}f_{2}^{\text{eff}}\bigg(g(\omega-\epsilon_1-\epsilon_2)+g(\omega)+\frac{1}{\omega+\epsilon_2-2\Delta+2i\Gamma}-\frac{1}{\omega-\epsilon_2-2\Delta+2i\Gamma}-2g(\omega+\epsilon_1-\epsilon_2)\bigg) \bigg].
\end{align}
It is simple to check that at the particle-hole symmetry point when either $\Phi=0$ or $\Delta T=0$ (or the equilibrium case of both) this reduces to the expression in Ref.\ \onlinecite{DuttLeHur}. 

We extract the proper self-energy by expanding Eqs.\ \eqref{eq:Dyson} and \eqref{eq:Dyson_proper} to second order
\begin{align}
G^{\text{ret}}_{d_{\sigma}d^{\dag}_{\sigma}}(\omega)=G^{\text{ret}(0)}_{d_{\sigma}d^{\dag}_{\sigma}}(\omega)+G^{\text{ret}(0)}_{d_{\sigma}d^{\dag}_{\sigma}}(\omega)\left(\Sigma^{(1)}(\omega)+\Sigma^{(2)}(\omega)\right)G^{\text{ret}(0)}_{d_{\sigma}d^{\dag}_{\sigma}}(\omega)
\label{eq:pert_dyson_1}
\end{align}
and
\begin{align}
G^{\text{ret}}_{d_{\sigma}d^{\dag}_{\sigma}}(\omega)=G^{\text{ret}(0)}_{d_{\sigma}d^{\dag}_{\sigma}}(\omega)+G^{\text{ret}(0)}_{d_{\sigma}d^{\dag}_{\sigma}}(\omega)\left(\Sigma^{\star(1)}(\omega)+\Sigma^{\star(2)}(\omega)\right)\left(G^{\text{ret}(0)}_{d_{\sigma}d^{\dag}_{\sigma}}(\omega)+G^{\text{ret}(0)}_{d_{\sigma}d^{\dag}_{\sigma}}(\omega)\Sigma^{\star(1)}(\omega)G^{\text{ret}(0)}_{d_{\sigma}d^{\dag}_{\sigma}}(\omega)\right)
\label{eq:pert_dyson_2}
\end{align}
Comparing Eqs.\ \eqref{eq:pert_dyson_1} abd \eqref{eq:pert_dyson_2} we find
\begin{align}
\Sigma^{\star(1)}(\omega)&=\Sigma^{(1)}(\omega)\notag\\
\Sigma^{\star(2)}(\omega)&=\Sigma^{(2)}(\omega)-g(\omega)\left(\Sigma^{(1)}(\omega)\right)^2
\end{align}
Thus we obtain
\begin{empheq}[box=\fbox]{align}
\Sigma^{\star(1)}=U\bigg[\frac{\Gamma}{\pi}\int d\epsilon\left| g(\epsilon)\right|^{2}f^{\text{eff}}(\epsilon)-\frac{1}{2}\bigg].
\label{eq:self_1st_proper}
\end{empheq} 
and similarly
\begin{empheq}[box=\fbox]{align}
&\Sigma^{\star(2)}=U^2\bigg[\frac{\Gamma}{\pi}\int d\epsilon \frac{ f^{\text{eff}}(\epsilon)\left|g(\epsilon)\right|^2}{\omega+\epsilon-2\Delta+2i\Gamma}+\left(\frac{\Gamma}{\pi}\right)^2\int d\epsilon_1 d\epsilon_2 \left|g_1\right|^2\left|g_2\right|^2 f_{1}^{\text{eff}}f_{2}^{\text{eff}}\bigg\{g(\omega-\epsilon_1-\epsilon_2)-2g(\omega+\epsilon_1-\epsilon_2)\bigg\}+\notag\\
&\left(\frac{\Gamma}{\pi}\int_{-\infty}^{\infty} d\epsilon \left|g(\epsilon)\right|^2 f^{\text{eff}}(\epsilon)-\frac{1}{2}\right)\left\{\frac{\Gamma}{\pi}\int d\epsilon' f^{\text{eff}}(\epsilon')\left(\frac{2\left(\epsilon'-\Delta\right)}{(\left(\epsilon'-\Delta\right)^2+\Gamma^2)^2}-\left(\frac{\left|g(\epsilon')\right|^2}{\omega-\epsilon'-2\Delta+2i\Gamma}-\frac{\left|g(\epsilon')\right|^2}{\omega-\epsilon'+2i\Gamma}\right)\right)\right\}\bigg].
\end{empheq}

\end{widetext}

\end{document}